%% file: main.tex
\documentclass[10pt,letterpaper,twocolumn]{article}

\usepackage[pagenumbers]{cvpr} 

\definecolor{cvprblue}{rgb}{0.21,0.49,0.74}
\usepackage[pagebackref,breaklinks,colorlinks,allcolors=cvprblue]{hyperref}
\usepackage{bm}
\usepackage{amsmath}
\usepackage{graphicx}
\usepackage[linesnumbered, ruled, lined]{algorithm2e}
\usepackage{multirow}
\usepackage{placeins}
\usepackage{utfsym}
\usepackage{cuted}      
\usepackage{caption} 
\usepackage{float}
\usepackage{xcolor}
\usepackage[most]{tcolorbox}
\usepackage{booktabs}
\usepackage{multirow}
\usepackage[table]{xcolor}

\usepackage{algorithmic}
\usepackage{amssymb}

\DeclareFixedFont{\ttb}{T1}{txtt}{bx}{n}{8} 
\DeclareFixedFont{\ttm}{T1}{txtt}{m}{n}{8}  

\usepackage{color}
\definecolor{deepblue}{rgb}{0,0,0.5}
\definecolor{deepred}{rgb}{0.6,0,0}
\definecolor{deepgreen}{rgb}{0,0.5,0}

\def\paperID{6148} 
\def\confName{CVPR}
\def\confYear{2025}

\tcbset{
  llmprompt/.style={
    colback=cyan!6,
    colframe=cyan!60,
    coltitle=black,
    fonttitle=\bfseries,
    boxrule=0.5pt,
    arc=2mm,
    top=1mm,
    bottom=1mm,
    left=2mm,
    right=2mm
  }
}

\title{Text2Villa: Hierarchical Generation of 3D Indoor Environments with Physics-Aware Analysis-by-Synthesis}

\author{
Xiang Tang$^{1,2}$ \quad
Ruotong Li$^{2}$ \quad
Xiaopeng Fan$^{3,2,4}$ \\
$^1$Harbin Institute of Technology, Shenzhen \quad $^2$Peng Cheng Laboratory \\ 
$^3$Harbin Institute of Technology \quad $^4$Harbin Institute of Technology, Suzhou Research Institute\\
}

\begin{document}

\twocolumn[{
\renewcommand\twocolumn[1][]{#1}
\maketitle

\input{pic/teaser}
}]

\input{sec/0_abstract}

\input{sec/1_introduction}
\input{sec/2_relatedwork}
\input{sec/3_method}
\input{sec/4_experiment}
\input{sec/5_discussion}
\input{sec/6_conclusion}

{
    \small
    \bibliographystyle{ieeenat_fullname}
    \bibliography{main}
}

\input{sec/7_appendix}


\end{document}

%% file: pic/teaser.tex

\begin{center}
    \centering
    \vspace{-0.3cm}
    \includegraphics[width=\textwidth]{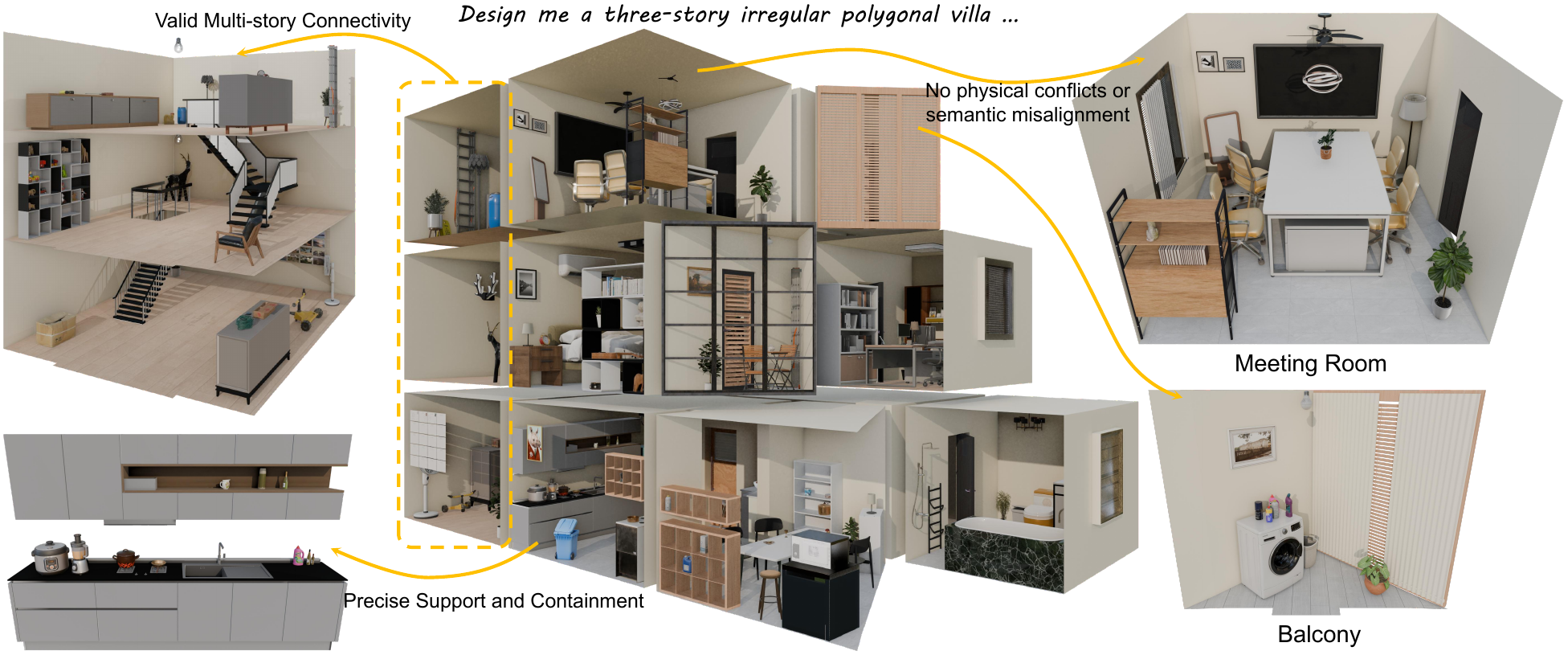} 
    
    \captionof{figure}{
        Text2Villa generates villa-scale 3D indoor environments from text. At the macro-level, our method produces diverse building foundations featuring multi-story connectivity, strictly adhering to user intent. At the micro-level, the fine-grained asset instantiation achieves precise physical support, cavity containment, and semantic alignment.
    }
    \label{teaser}
\end{center}
\vspace{0.4cm} 

%% file: sec/0_abstract.tex
\begin{abstract}
Generating 3D indoor scenes from natural language holds tremendous potential, yet existing methods predominantly fail to generate multi-room structures with vertical connectivity and arbitrary polygonal boundaries. Furthermore, they lack a deep grounding in continuous 3D physical laws, leading to severe geometric penetrations and floating artifacts. In this work, we propose Text2Villa, a novel hierarchical generative framework. At the macro level, we construct a multi-story dataset to fine-tune an autoregressive layout generator, ensuring the direct parsing of text into 3D building foundations featuring polygonal boundaries and multi-story connectivity. To enforce physical laws during micro-level asset arrangement, we introduce the Affordance-driven Physical-Semantic Scene Graph (A-PSSG) to explicitly abstract physical affordances (such as support surfaces and containment cavities) into node attributes, establishing strict geometric and semantic edge constraints. Guided by the A-PSSG, we formulate scene instantiation as a constrained closed-loop optimization problem following the analysis-by-synthesis paradigm. By integrating an underlying geometric collision detection engine with the high-level semantic reasoning of multimodal large language models (MLLMs), our heuristic solver dynamically executes physics-aware actions under the observation-evaluation-modification mechanism to effectively resolve mesh collisions, floating artifacts, and fine-grained cavity containment failures. Extensive experiments demonstrate that Text2Villa outperforms previous methods across various metrics, robustly generating high-fidelity and physically plausible villa-level 3D environments from text, thereby providing a reliable and interactive 3D content foundation for downstream applications.

\end{abstract}

%% file: sec/1_introduction.tex
\section{Introduction}

Automatically generating high-fidelity and logical 3D indoor scenes holds immense application potential in virtual reality, game development, interior design, and embodied AI training. These downstream applications require generated environments with rational spatial layouts and physically reliable interactions. Moreover, populating large-scale scenes necessitates an abundant and diverse supply of 3D assets. Driven by the rapid development of MLLMs and generative artificial intelligence, natural language guided 3D scene generation \cite{yang2024holodeck, zhou2024gala3d, ccelen2024design, feng2025text} draws widespread attention. While recent pioneering works \cite{hu2024scenecraft, ling2025scenethesis, li2025phip, yang2025sceneweaver, luo2026sceneassistant} design autonomous agents that significantly improve generation efficiency and facilitate large-scale scene construction through multi-turn reasoning, they still face a massive gap when processing highly complex real-world indoor environments.

Most mainstream indoor scene generative methods \cite{sun2025layoutvlm, gu2025artiscene} are limited to macroscopic assumptions of single rooms and typically predefine room boundaries as simple rectangular contours. Buildings in the real world, however, are inherently holistic, featuring multi-story structures with diverse exterior footprints and strict vertical constraints. Naively stacking single-story layouts generated by existing methods inevitably fails, as it ignores crucial inter-floor dependencies such as precise staircase alignments and coherent exterior boundaries. The method most closely related to ours is MANSION \cite{che2026mansion}, which establishes room connections via bubble diagrams, achieves multi-floor connectivity using vertical core structures, and ultimately instantiates the layouts into 3D scenes within the AI2-THOR simulator. However, this approach remains unable to generate irregular building and room footprints with edges at arbitrary angles. Regarding room-level micro-arrangement, existing feed-forward generation models \cite{yang2024physcene, tang2024diffuscene} generally lack a deep grounding of 3D continuous physical space. This deficiency frequently leads to floating objects, severe geometric collisions, or irrational placements.

Moreover, in practical digital scene development, synthesizing these complex architectures requires bridging the gap between massive 3D asset acquisition and rigorous physical-semantic verification. Traditional pipelines struggle to efficiently source appropriate models and lack a continuous feedback mechanism to correct microscopic placement errors, which severely undermines the physical realism and usability of the generated scenes. To address these bottlenecks and emulate the top-down, physically and functionally bounded human design process, we propose Text2Villa, a hierarchical generation framework that decouples macro-level architectural structure derivation from micro-level scene instantiation.

\input{pic/pipeline}

First, we construct and validate a dataset specifically for multi-story building generation which precisely parameterizes spatial boundaries using geometric polygon primitives and injects connectivity and usability priors. Through autoregressive fine-tuning on this dataset, our architectural layout generator learns implicit architectural rules and robustly outputs globally consistent 3D multi-story building foundations. Second, we introduce a nested graph to represent the scene, in which each room node in the spatial layout graph is recursively expanded into our proposed A-PSSG for asset arrangement. It goes beyond traditional spatial scene graphs by explicitly abstracting the physical affordances of objects as multiple attributes of graph nodes and strict geometric intersection rules as edge constraints. Based on A-PSSG, we formulate the scene instantiation problem as a constrained closed-loop iterative optimization process following the analysis-by-synthesis concept. Our heuristic solver combines a physical engine with the visual semantic feedback of MLLMs to dynamically perform physics-aware actions within a continuous observe-evaluate-modify closed-loop search, which can not only automatically correct the semantic positional relationships among furniture but also successfully resolves fine-grained mesh penetrations, support failures, and cavity containment problems within restricted parameter spaces. The contributions of this work are summarized as follows:

\begin{itemize}
\item We propose a hierarchical generation framework, Text2Villa, capable of directly parsing text into multi-story 3D building structures. This macro-level capability is powered by an autoregressive layout generator fine-tuned on a custom-built dataset with architectural usability constraints.
\item We propose a nested graph to hierarchically model 3D scenes, involving a layout graph for macro-level spatial configuration and an A-PSSG for micro-level asset placement. The A-PSSG abstracts multiple functional affordances like support surfaces and containment cavities into node attributes and further constructs dual physical-semantic edge constraints accordingly, providing a robust theoretical framework for the physically realistic generation of 3D scenes.
\item Based on the A-PSSG, we design an analysis-by-synthesis instantiation solver to ensure that objects acquired from diverse asset sources are physically plausible within the spaces. By integrating geometric hard constraint verification and MLLM heuristic evaluation, this solver effectively addresses mesh collisions, floating, and semantic errors in fine-grained asset configuration, thereby achieving high-fidelity indoor scene synthesis.
\end{itemize}

%% file: pic/pipeline.tex
\begin{figure*}
    \centering
    \includegraphics[width=\textwidth]{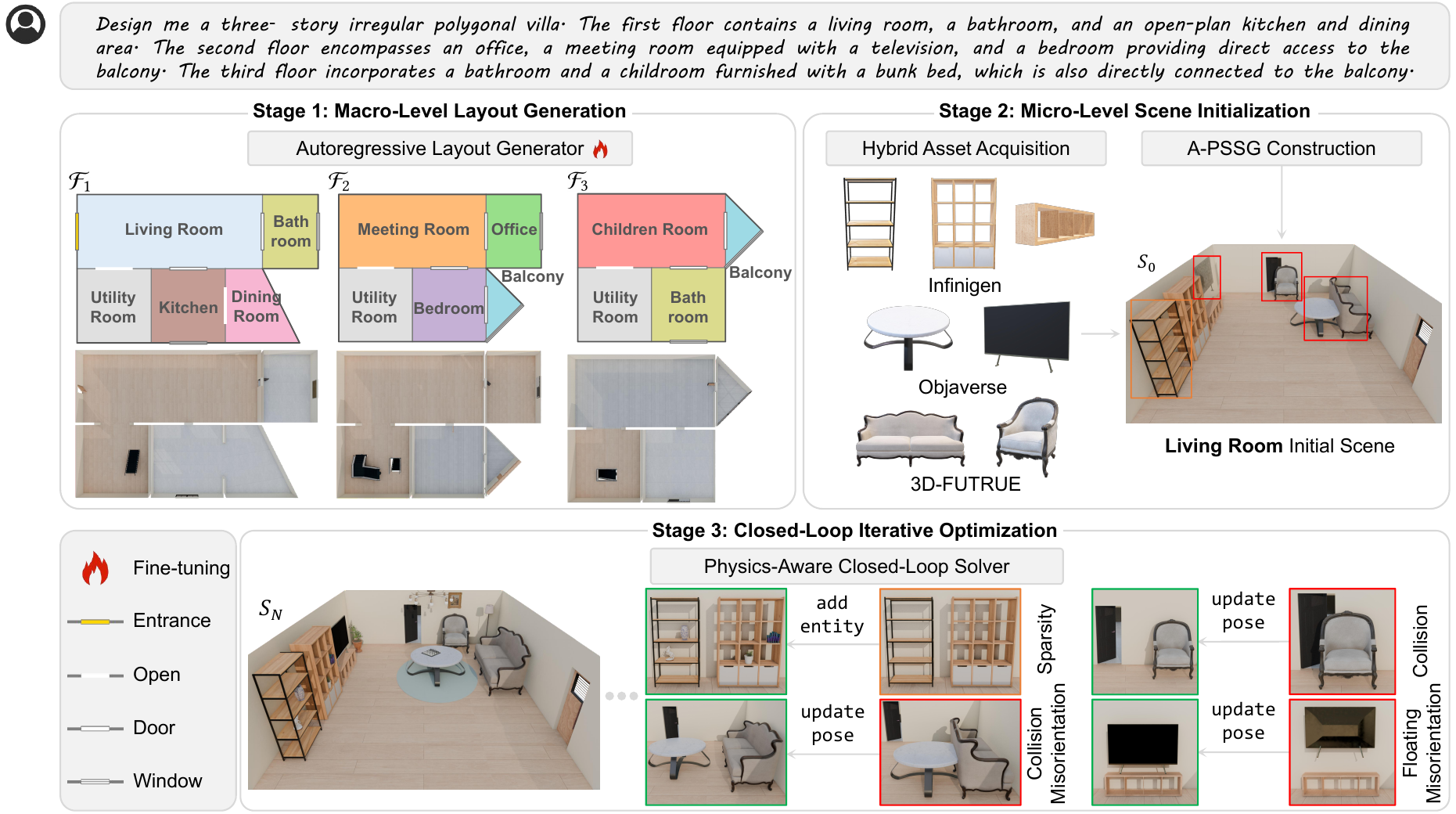}
    \caption{Pipeline bridging macro building generation and micro object planning. Stage 1: A fine-tuned autoregressive layout generator parses user prompts into parameterized JSON sequences, and its geometric parsing module then converts into a 3D multi-story building. Stage 2: For each generated room node (e.g., living room), we construct an A-PSSG based on 3D asset categories inferred by an MLLM to establish physical and semantic constraints. Combined with a hybrid asset acquisition, this outputs an initial, coarsely arranged scene configuration $S_0$ within the building container. Stage 3: Initialized with $S_0$, a physics-aware heuristic solver executes closed-loop iterative optimization. Through a continuous observation-evaluation-modification cycle, it progressively resolves physical conflicts and semantic errors, ultimately producing a high-fidelity, physically plausible scene layout $S_N$.}
    \label{pipeline}
\end{figure*}

%% file: sec/2_relatedwork.tex
\section{Related work}

\paragraph{Text-driven 3D Scene Generation} Previous text-driven 3D scene generation primarily relies on predefined rules \cite{raistrick2024infinigen, gumin2025procedural} or graph-based retrieval \cite{wang2018deep}. Recently, autoregressive \cite{paschalidou2021atiss, zhao2024roomdesigner} and diffusion models \cite{lin2024instructscene, bokhovkin2025scenefactor, meng2025lt3sd, li2026worldgrow} drive significant progress in this field. Diffusion-based methods typically generate panoramas or meshes through depth estimation and image inpainting. Recent advancements \cite{zhang2025generating, yang2025layerpano3d} further push the boundaries of visual consistency through hyper-immersive scene generation and wide-baseline panorama reconstruction. While producing visually coherent results, these approaches lack independent 3D asset entities and thus hinder subsequent physical interactions. Other approaches based on asset retrieval and arrangement \cite{ocal2024sceneteller, yang2025sceneweaver, hao2025mesatask, gu2025artiscene} generate scenes comprising independent 3D models. Most of these methods restrict their scope to a single room with preset rectangular boundaries. In contrast, Text2Villa directly parses text to generate complex villa-level architectural structures featuring polygonal boundaries and multi-story connectivity.

\paragraph{Scene Graph Representation for Layouts} Scene graphs serve as a structured representation widely applied in 3D scene understanding and synthesis. In existing layout generation research \cite{gao2024graphdreamer, yang2025mmgdreamer, zhu2025imaginarium, zhou2026layoutdreamer}, scene graphs primarily capture the relative semantic spatial relationships among nodes. Some studies \cite{zhang2025open, wang2026function2scene, hu2026hierarchical} have extended these into functional 3D scene graphs to guide spatial layouts based on functional design constraints. Real interior design follows strict physical laws where spatial semantics alone fail to accurately guide micro-level arrangements, such as placing a book inside a specific shelving unit. While the concept of affordance \cite{jiang2024affordance, chu20263d} is crucial for resolving these physical interactions, it has not yet been integrated into scene graph representations. The breakthrough of this work lies in the proposal of A-PSSG. It explicitly abstracts physical attributes like support surfaces and containment cavities into internal features of graph nodes, and establishes dual physical and semantic constraints for the edges accordingly, providing a mathematical foundation for precise geometric intersection and placement.

\paragraph{Layout Generation and Optimization in Spatial Reasoning} In the 2D domain, floorplan generation has seen notable advancements via centroid-guided diffusion \cite{lian2026cg} and markup representations \cite{Shiohara2026}. Lifting to 3D space, scene layout generation constitutes a combinatorial optimization problem requiring spatial reasoning and geometric constraints. Existing methods combine decomposition-recomposition \cite{chen2024comboverse, han2025reparo, wu2025diorama}, depth/point cloud alignment \cite{zhou2024zero, Ardelean2025Gen3DSR, yao2025cast, tang2026towards, tang2025zeroscene}, and multi-instance attention \cite{huang2025midi, meng2026scenegen} to update multi-asset poses and restore scene layouts. Meanwhile, large language models (LLMs) demonstrate remarkable capabilities in commonsense reasoning and zero-shot task planning, prompting several works \cite{feng2023layoutgpt, li2024dreamscene, sun2025hierarchically} to employ them for layout generation. Inherently trained on text corpora, LLMs lack a genuine perception of continuous 3D physical spaces. Consequently, layouts generated in a feed-forward manner appear semantically reasonable at a macroscopic level while containing extensive physical hallucinations at a microscopic level. Overcoming this limitation, recent studies \cite{che2026mansion, pfaff2026scenesmith, xia2026sage} have shifted towards agentic frameworks, aiming to generate simulation-ready scenes for embodied AI. Aligning with this trend, and to achieve more precise layout generation, we integrate semantic spatial reasoning with physical optimization and propose an analysis-by-synthesis closed-loop iterative instantiation solver. We utilize a physics engine to calculate support and collision penalties, and an MLLM to assess scene semantics. Within continuous iterations of observation, evaluation, and modification, our solver guides pose search, optimizes layouts, and provides semantic feedback.

%% file: sec/3_method.tex
\section{Text2Villa}

Text2Villa takes unstructured natural language describing complex buildings as input and outputs fully instantiated, physically plausible 3D indoor scenes. We formulate the scene generation as a nested generative process, as illustrated in Fig. \ref{pipeline}.  At the macro level, we model the building as a structural layout graph for the parametric representation of floor (Subsec. \ref{subsec3.1}). At the micro level, each room node within this macro graph acts as a parent container, recursively expanding into our proposed A-PSSG (Subsec. \ref{subsec3.2}) to define fine-grained object relations. Following this nested graph construction, the pipeline proceeds with hybrid 3D asset acquisition (Subsec. \ref{subsec3.3}) and a physics-aware closed-loop iterative instantiation (Subsec. \ref{subsec3.4}).

\subsection{Multi-Story Architectural Layout Generation}\label{subsec3.1}

Mapping text into complex 3D architectural structures is the initial step of the scene generation pipeline. We model the architectural layout as a hierarchical data composition formally represented as $\mathcal{H} = \{\mathcal{F}_1, \dots, \mathcal{F}_N\}$, where $\mathcal{F}_i$ represents the $i$-th floor. Each floor $\mathcal{F}_i$ is viewed as a spatial layout graph $(\mathcal{R}, \mathcal{E})$. The nodes $\mathcal{R}$ represent various room instances covering ten core functional areas (such as living rooms, bedrooms, and kitchens) with polygonal boundaries. The edges $\mathcal{E}$ explicitly encode the connectivity between rooms. These relationships include restricted connections with physical partitions such as \texttt{entrances}, \texttt{doors}, and \texttt{windows}, \texttt{open} connections, and multi-story vertical connections such as \texttt{staircases} located in utility rooms. This macro-level spatial graph serves as a basis for the micro-level graph expansion.

\input{pic/graph}

Based on the above definition, we construct and validate a fine-tuning dataset dedicated to multi-story building generation comprising 1,000 high-quality text-to-architecture JSON pairs (additional details are provided in Appendix \ref{app:dataset}). At the data representation level, we parameterize spatial boundaries and wall segments using geometric primitives. This allows us to define various exterior architectural contours including H-shape, L-shape, T-shape, U-shape, and irregular polygons while ensuring the internal room layout supports arbitrary polygonal shapes. We introduce rationality prior constraints based on real-world architectural design during dataset construction. For instance, all rooms must be accessible, bathrooms and kitchens require windows, and balconies can feature panoramic windows. Using this dataset, we fine-tune an LLM-based \cite{qwen3.5} autoregressive layout generator. By parsing the input text, the generator outputs structured JSON sequences, and its geometric parsing module converts the parametric information into multi-story building volume containing 3D base walls, floors, and door or window meshes. This process provides a precise physical space for subsequent fine-grained 3D asset population.

\subsection{Affordance-driven Physical-Semantic Scene Graph}\label{subsec3.2}

Our Text2Villa represents the scene as a nested graph, capturing the macro-level space configuration via a layout graph and encoding the object arrangement with an A-PSSG, as illustrated in Fig. \ref{graph}. Formally defined as $\mathcal{G} = (\mathcal{V}, \mathcal{E}_{sem}, \mathcal{E}_{phys})$, the A-PSSG introduces explicit physical affordances and strict geometric constraints to resolve physical ambiguities. The node and edge representations are defined as follows:

\paragraph{Node physical affordance}

Unlike previous hard partition methods based on object size or a single semantic category, we abstract the physical affordance of objects as multi-assignable graph node attributes. A node in the graph dynamically assumes three core interactive roles. 

\begin{itemize}
    \item Support providers $\mathcal{V}_{pro}$ such as floors, tables, or cabinets expose valid support surfaces $\mathcal{S}$ outwards or provide collision-free containment cavities $\mathcal{C}$ inwards.
    \item Support consumers $\mathcal{V}_{con}$ such as desk lamps, books, or murals are constrained by gravity or attachment dependencies and must establish a physical connection with the support surfaces $\mathcal{S}$ or cavities $\mathcal{C}$ provided by $\mathcal{V}_{pro}$. 
    \item Functional interactors $\mathcal{V}_{int}$ such as sofas or wardrobes define an interaction reservation zone along their functional orientation to allocate space for human activities or part movements.
\end{itemize}

\paragraph{Edge dual constraints} In A-PSSG, edges describe relative positions and introduce geometric and physical constraints. 

\begin{itemize}
    \item Semantic relation edges $\mathcal{E}_{sem}$ define the relative orientation between objects. For instance, $(O_i, \text{face\_to}, O_j)$ requires the angle between their functional orientation normal vectors to fall within a specific threshold. 
    \item Physical interaction edges $\mathcal{E}_{phys}$ include support relations $(O_i, \text{ontop}, O_j)$ and containment relations $(O_i, \text{inside}, O_j)$. Specifically, the former requires the bottom bounding box of $O_i$ to be coplanar with the support surface $\mathcal{S}$ of $O_j$ and projected inside it. The latter further requires the geometry of $O_i$ to have no collision with the solid part of $O_j$ and to be located within an independent cavity $\mathcal{C}$ of $O_j$, provided the support condition is met.
\end{itemize}

\input{pic/algorithm}

In order to initiate the A-PSSG efficiently, we conduct hierarchical instructions to fully utilize the reasoning capability of MLLMs \cite{hurst2024gpt}. Concretely, we precisely instantiate the objects explicitly mentioned in the text input into a node set $\mathcal{V}$, corresponding to the asset categories $C = \{c_1, c_2, \ldots, c_n\}$. Based on the spatial logic parsed from the text and implicit real-world object associations, these nodes are assigned the aforementioned physical affordance attributes, thereby establishing semantic relation edges $\mathcal{E}_{sem}$ among them. Meanwhile, physical interaction edges $\mathcal{E}_{phys}$ are built between attribute ports $\mathcal{V}_{pro}$ and $\mathcal{V}_{con}$ based on affordance matching. If the user input is brief, the MLLM leverages common-sense priors to supplement reasonable asset categories and quantities according to the room type $\mathcal{R}$, and connects the edges (the system prompt for constructing the A-PSSG is detailed in Appendix Table \ref{prompt2}). This process successfully translates unstructured text into physical and semantic contracts that guide subsequent 3D spatial layout solving.

\subsection{Hybrid 3D Asset Acquisition}\label{subsec3.3}

To balance scene diversity, semantic alignment, and the precise geometric requirements of specific furniture, we adopt a hybrid asset acquisition. This strategy encompasses procedural generation, cross-modal retrieval, and style-aware extraction, ensuring both high-quality geometric interactions and open-vocabulary richness.

For structural categories $c_i \in C_{struc}$ (e.g., shelving units, cabinets), geometric accuracy is paramount for determining affordances. Here, we prioritize procedural generation frameworks like Infinigen \cite{raistrick2024infinigen}. Beyond merely generating the solid mesh, Text2Villa explicitly computes and exports a set of available support surfaces $\mathcal{S}$ and a set of internal free-space bounding boxes $\mathcal{C}$ during generation. For other objects, we involve a retrieval method that can query any large-scale 3D dataset, as detailed in our acquisition pipeline (see Algorithm \ref{alg:hybrid_acquisition}). We use the text category of the graph node as a query condition and extract the text embedding and compute the cosine similarity against the 3D assets within a unified shape latent space \cite{liu2023openshape}. Then, we conduct boolean checks using Trimesh (a Python library for 3D mesh processing) to filter out models that are non-watertight, self-intersecting, or possess abnormal face counts, sort the remaining candidates in descending order by their similarity scores, and extract the best-matching mesh model. Furthermore, by integrating metadata tags and asset texture editing methods \cite{tang2025zeroscene}, our hybrid asset acquisition also exhibits potential in stylized scene generation when user prompts specify distinct stylistic preferences (e.g., a rustic medieval tavern or a vintage detective’s study, as shown in Fig. \ref{stylize}).

\input{pic/stylize}

Semantic relation edges $\mathcal{E}_{sem}$ in the A-PSSG depend on exact knowledge of canonical object orientations. Thus, we introduce a preprocessing step before asset storage. We render four orthogonal views for each retrieved 3D model around the $Z$-axis at intervals of $90^\circ$ and input them into the MLLM alongside specific prompts, requiring it to identify the image representing the functional front of the object. The extracted forward offset angle is recorded in the asset metadata, laying the foundation for the closed-loop solver in Subsec. \ref{subsec3.4} to execute precise pose optimization.

\subsection{Physics-Aware Closed-Loop Solver}\label{subsec3.4}

\input{pic/solver}

We formulate the A-PSSG instantiation process as an optimization problem with geometric constraints to determine the exact six-degree-of-freedom pose $(T, R)$ and scale $D$ for each 3D asset in the scene. Fig. \ref{solver} illustrates this iterative resolution process through a concrete scene instantiation example. We adopt an analysis-by-synthesis paradigm, which leverages an MLLM and a physics engine collaboratively to minimize the overall energy of the generated scene $E(S_t)$, defined as:
\begin{equation}
\label{eq1}
\resizebox{0.9\linewidth}{!}{$ \displaystyle
    E(S_t) = \lambda_{col} E_{col}(S_t) + \lambda_{sup} E_{sup}(S_t) + \lambda_{sem} E_{sem}(S_t)
$}
\end{equation}
where $S_t$ denotes the scene state at the $t$-th iteration of our optimization process. Based on the A-PSSG, we compute the following energy terms: the semantic penalty $E_{sem}$ is evaluated by the MLLM. Taking multi-view rendered images of the current scene as input, it evaluates whether the relative object relationships defined by $\mathcal{E}_{sem}$ are satisfied, whether appropriate interaction zones are reserved for nodes with the $\mathcal{V}_{int}$ attribute, and assesses the overall richness of the scene. Meanwhile, $E_{col}$ and $E_{sup}$ are calculated via the physics engine. Specifically, $E_{col} = \sum_{i \neq j} \text{Vol}(O_i \cap O_j)$ is the collision penalty that calculates the intersection volume between any pair of independent objects $O_i$ and $O_j$ in the scene. This term drops to zero if and only if the two objects do not overlap. Similarly, the support penalty is defined as the sum of the distance error and the projection out-of-bounds error:
\begin{equation}
\label{eq2}
\resizebox{0.9\linewidth}{!}{$ \displaystyle
    E_{sup} = \sum_{(i,j)\in \mathcal{E}_{phys}} \big( \| p_{bottom}^i - p_{proj}^i \|_2 + \mathbb{I}(p_{proj}^i \notin \mathcal{S}_j) \cdot \text{Penalty} \big)
$}
\end{equation}
where $p_{bottom}^i$ represents the bottom center point of object $O_i$, and $p_{proj}^{i}$ denotes its projection onto the plane containing the valid support surface $\mathcal{S}_j$ of the provider object $O_j$, cast along the negative normal direction of $\mathcal{S}_j$. $\mathbb{I}$ is an indicator function. This penalty is evaluated only for nodes with the $\mathcal{V}_{con}$ attribute to ensure gravity and attachment constraints.

\input{pic/optimization}

\paragraph{Analyzing-Updating Loop} The instantiation process begins at the initial state $S_0$, where we roughly initialize the base asset set $\mathcal{A}$ into the architectural container (generated in Sec. \ref{subsec3.1}) based on the text prompts and the constraints defined in the A-PSSG. Because the scene at this step typically involves geometric conflicts and layout blanks, the solver determines the optimization direction by computing the aforementioned penalties, continuously generating new A-PSSGs and states $S_{t+1}$ through iteration. During each update step $t$, the physics engine receives discrete atomic actions output by the MLLM (including \texttt{add\_entity} and \texttt{remove\_entity} to improve scene completeness and resolve irreconcilable collisions; \texttt{update\_pose} to output a translation $\Delta T$ and rotation $\Delta R$, together with \texttt{update\_scale} to adjust the placements and dimensions of the assets), and conducts rigorous validation in the continuous parameter space to further reduce the penalty terms. The optimization stops when the total scene energy $E(S_t)$ converges or drops below a predefined safety threshold.

\paragraph{Fine-grained resolving for inside constraints} Handling the constraint $(O_i, \text{inside}, O_j)$ requires specific treatment during the update loop. Traditional collision detection triggers false alarms here since the bounding box of the parent object inevitably intersects with the child. Therefore, the optimization target specifically enforces that the bounding box $B_{child}$ of $O_i$ must be completely contained within a candidate cavity $\mathcal{C}_k \in \mathcal{C}$ of $O_j$, while their solid meshes must not intersect. Crucially, an internal relationship inherently dictates a valid support dependency; the child object cannot float and must rest firmly on the bottom surface of the cavity $\mathcal{C}_k$. As shown in Fig. \ref{optimization}, the solver enters a local search mode by triggering the \texttt{add\_entity} inside action. Guided by boundary detection and support verification from the physics engine, it continuously utilizes \texttt{update\_pose} to search the valid parameter space. Once a pose satisfying cavity containment, collision-free, and resting conditions is found, both the $E_{col}$ and $E_{sup}$ penalties for this inside constraint immediately drop to zero, thereby successfully locking in the precise microscopic asset placement.

%% file: pic/graph.tex
\begin{figure}[t]
    \centering
    \includegraphics[width=\columnwidth]{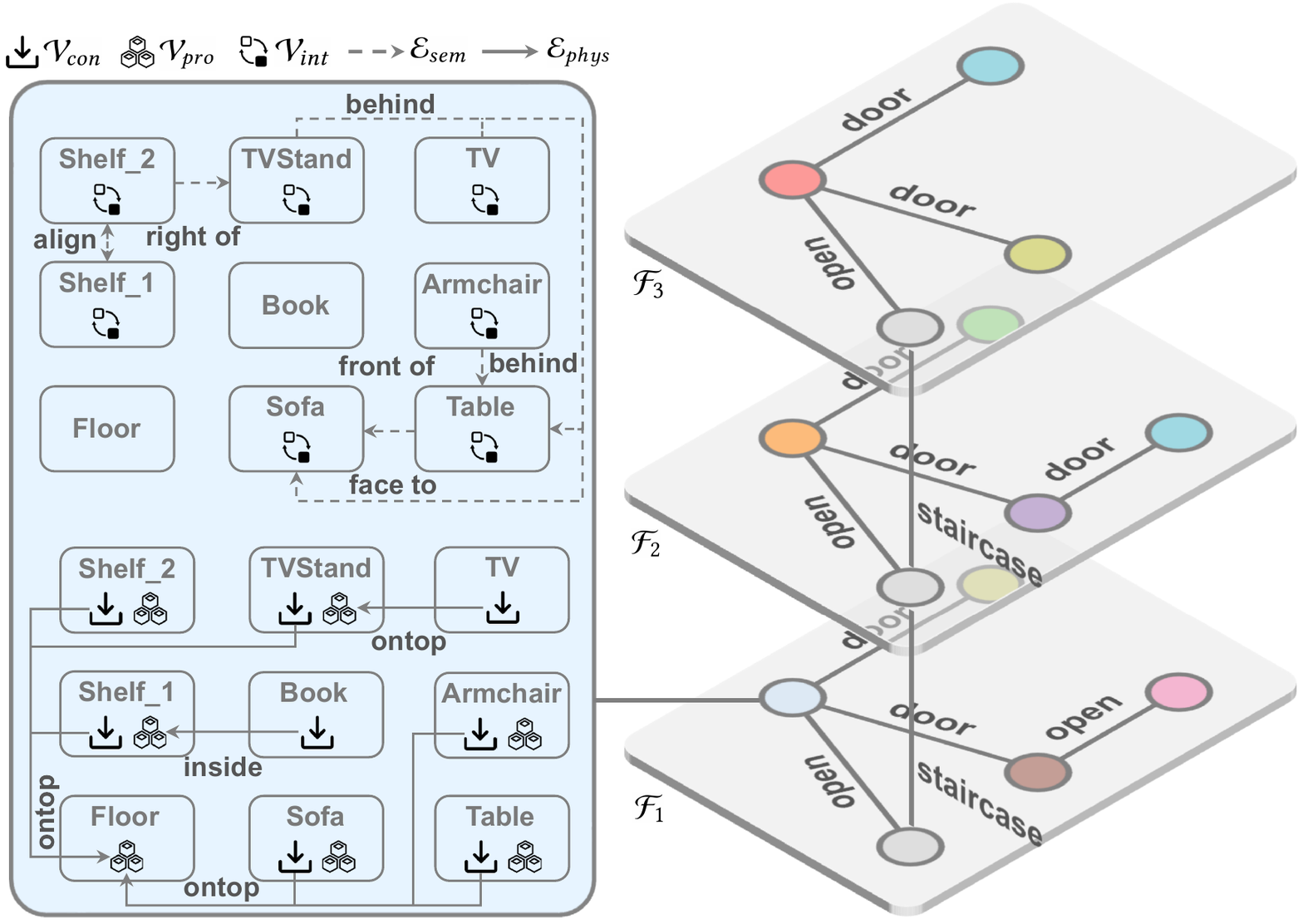}
    \vspace{-0.6cm}
    \caption{Nested graph representation for scenes. Text2Villa models the 3D environment hierarchically. Right: The spatial graph represents the macro-level architectural layout. Left: Each room node acts as a parent container that recursively expands into a micro-level A-PSSG (e.g., Living Room).}
    \vspace{-0.4cm}
    \label{graph}
\end{figure}

%% file: pic/algorithm.tex
\begin{algorithm}[t]
\SetAlgoNoLine
\KwIn{Target category set $C = \{c_1, c_2, \dots, c_n\}$, Structural categories $\mathcal{C}_{struc}$, 3D database $\mathcal{D} = \{(\mathbf{M}_j, \mathbf{f}_{j})\}_{j=1}^N$, Text Encoder $\mathcal{E}_{txt}$, Similarity threshold $\tau$, Candidate size $K$.}
\KwOut{Retrieved 3D asset mapping $\mathcal{A}$.}

Initialize asset dictionary $\mathcal{A} \leftarrow \emptyset$\;
\For{each target category $c_i \in C$}{

    \If{$c_i \in \mathcal{C}_{struc}$}{
        $\mathbf{M}^* \leftarrow \text{InfinigenGenerator}(c_i)$\;
        $\mathcal{A}[c_i] \leftarrow \mathbf{M}^*$\;
        \textbf{continue}\;
    }
    Construct prompt string $T_i \leftarrow \text{PromptTemplate}(c_i)$\;
    Extract text embedding $\mathbf{q}_i \leftarrow \mathcal{E}_{txt}(T_i)$\;
    Normalize query vector: $\hat{\mathbf{q}}_i \leftarrow \frac{\mathbf{q}_i}{\|\mathbf{q}_i\|_2}$\;
    
    Initialize candidate set $S_i \leftarrow \emptyset$\;
    \For{each 3D model $(\mathbf{M}_j, \mathbf{f}_{j}) \in \mathcal{D}$}{
        $s_{i,j} \leftarrow \hat{\mathbf{q}}_i \cdot \mathbf{f}_{j}$\;  
        \If{$s_{i,j} > \tau$ \textbf{and} $\text{ComplexityFilter}(\mathbf{M}_j)$ is True}{
            $S_i \leftarrow S_i \cup \{(\mathbf{M}_j, s_{i,j})\}$\;
        }
    }
    
    Sort $S_i$ in descending order of $s_{i,j}$ and keep Top-$K$ models\;
    $\mathbf{M}^* \leftarrow \text{Top-1}(S_i)$\;
    $\mathcal{A}[c_i] \leftarrow \mathbf{M}^*$\;
}
\KwRet $\mathcal{A}$\;

\caption{Hybrid 3D Asset Acquisition}
\label{alg:hybrid_acquisition}
\end{algorithm}

%% file: pic/stylize.tex
\begin{figure}[t]
    \centering
    \includegraphics[width=\columnwidth]{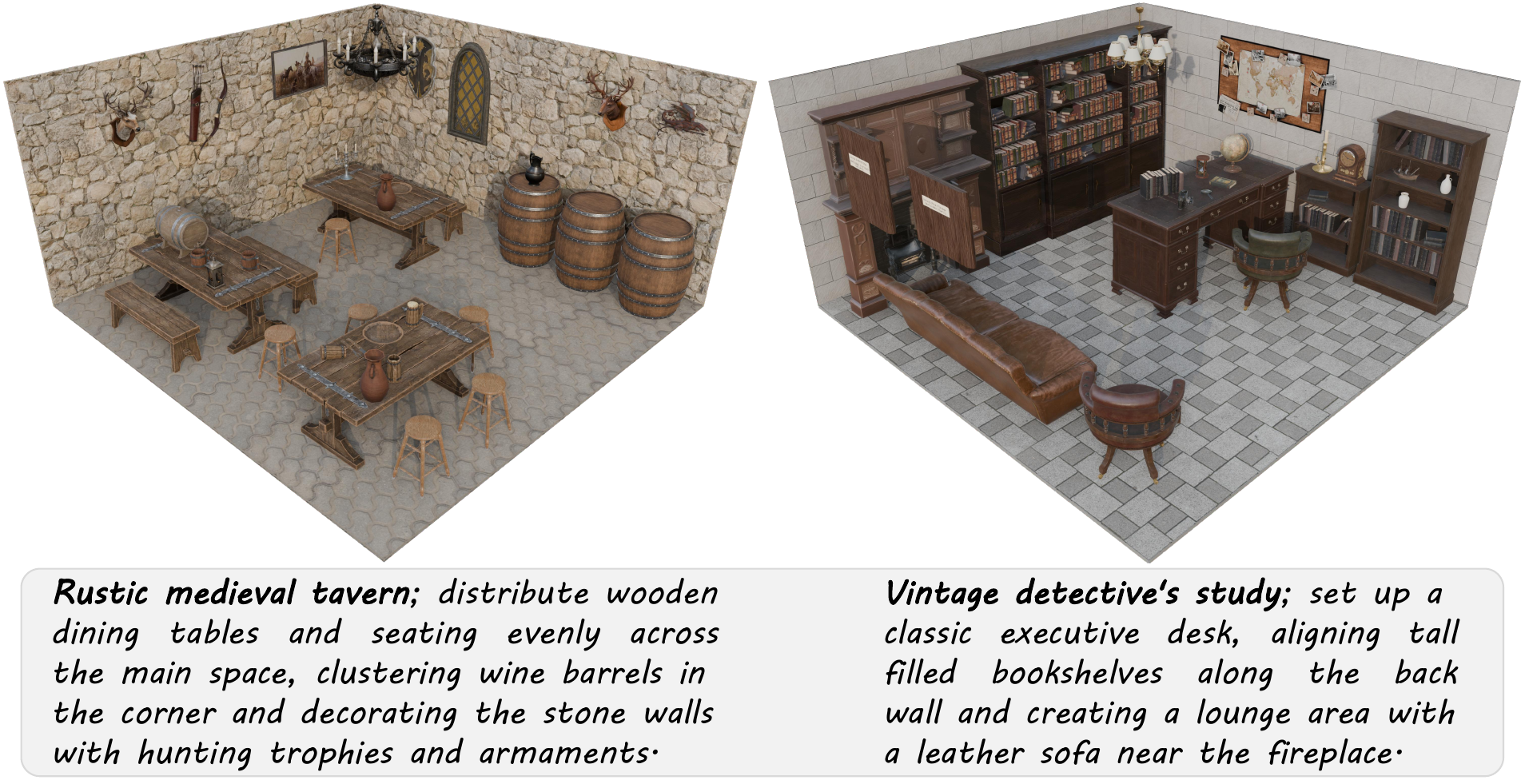}
    \vspace{-0.6cm}
    \caption{Stylized scene generated by Text2Villa.}
    \vspace{-0.4cm}
    \label{stylize}
\end{figure}

%% file: pic/solver.tex
\begin{figure}
    \centering
    \includegraphics[width=\columnwidth]{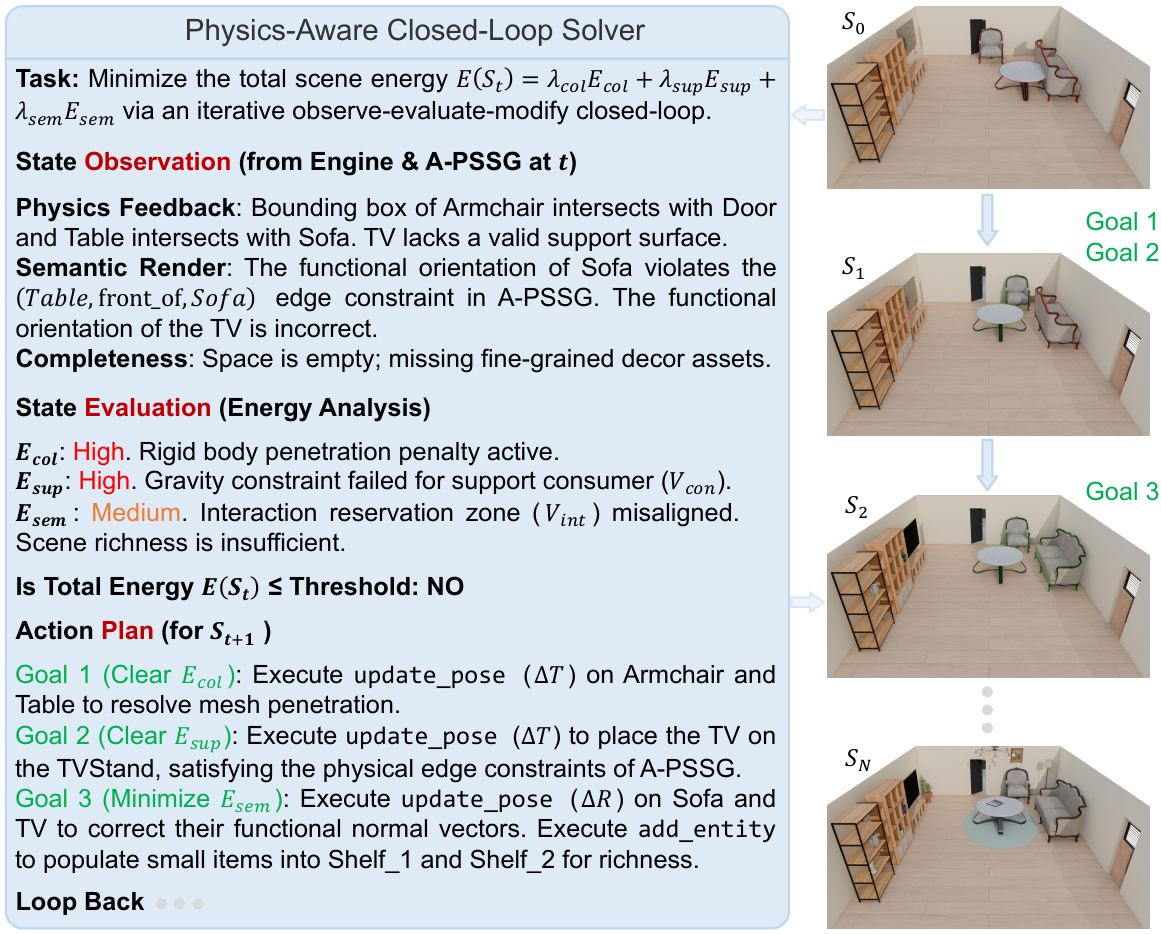}
    \vspace{-0.6cm}
    \caption{Visualization of the Physics-Aware Closed-Loop Solver. Following the analysis-by-synthesis paradigm, the solver takes observations from the physics engine and A-PSSG feedback to identify physical and semantic conflicts, evaluates the overall scene energy, and formulates an action plan to continuously iterate and optimize the scene configuration.}
    \vspace{-0.4cm}
    \label{solver}
\end{figure}

%% file: pic/optimization.tex
\begin{figure*}
    \centering
    \vspace{-0.2cm}
    \includegraphics[width=\textwidth]{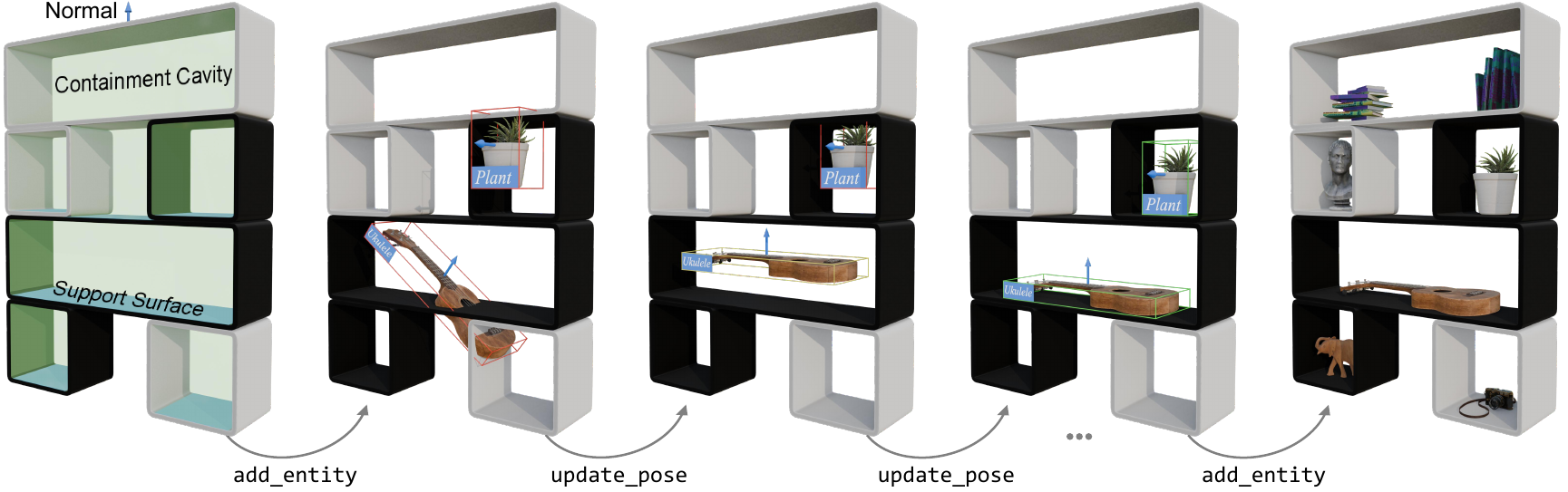}
    \vspace{-0.6cm}
    \caption{Fine-grained resolving process for inside constraints in the A-PSSG. The red bounding box indicates the solid mesh intersection between the child and parent objects. The yellow box denotes that the child object is successfully enclosed within a containment cavity of the parent. The green box represents that the solver has found for valid pose parameters, and the child object is accurately placed on the support surface inside the cavity of the parent object.}
    \label{optimization}
    \vspace{-0.4cm}
\end{figure*}

%% file: sec/4_experiment.tex
\section{Experiments}

\subsection{Experimental Setup}\label{subsec4.1}

\paragraph{Implementation Details} Our autoregressive layout generator builds upon Qwen3.5-9B \cite{qwen3.5} and undergoes fine-tuning with the LoRA strategy. We train the model for 15 epochs on a single NVIDIA L20 GPU using the AdamW optimizer with a batch size of 4, and a learning rate of 1e-4. For hybrid 3D asset acquisition, we integrate the Objaverse \cite{deitke2023objaverse} and 3D-FUTURE \cite{fu20213d} datasets to support cross-modal retrieval and utilize the geometry nodes of Infinigen \cite{raistrick2024infinigen} to generate structural assets. The MLLM utilized throughout the pipeline is GPT-4o \cite{hurst2024gpt}, and the physics engine is built upon the PyBullet and Trimesh libraries. In Eq. \ref{eq1}, the penalty weights are empirically set to $\lambda_{sem} = 1.0$, $\lambda_{col} = 5.0$, and $\lambda_{sup} = 2.0$.

\input{pic/qualitative0}

\paragraph{Baselines} We select state-of-the-art text-to-3D indoor scene generation methods based on LLMs or agents as primary baselines. The evaluation comprises two groups according to the task hierarchy. For macro-level architectural layout generation, we compare with MANSION \cite{che2026mansion}, Holodeck \cite{yang2024holodeck} and SceneFoundry \cite{chen2026scenefoundry}. Given that MANSION supports multi-story generation, we directly compare the generation quality of its overall multi-floor layouts. Since the other two methods inherently lack support for multiple floors, we relax the constraints and compare their multi-room generation capabilities with our single-story degraded results. For micro-level asset instantiation, we evaluate against Holodeck, MANSION, LayoutVLM \cite{sun2025layoutvlm}, and SceneWeaver \cite{yang2025sceneweaver}. These approaches focus on single-room generation and typically restrict room outlines to rectangles. To ensure a fair comparison, all methods use GPT-4o as the MLLM, and we keep the number and area of rooms consistent in macro-layout and single-room generation tasks respectively.

\input{pic/quantitative1}

\input{pic/qualitative1}

\paragraph{Evaluation Metrics} We customize several evaluation metrics for this task to measure multi-story connectivity and fine-grained physical interactions, with all implementation details elaborated in Appendix \ref{app:metrics}. At the macro-layout level, we measure Room Reachability (RR) to detect rooms lacking door connections or completely enclosed in the generated structure. We also use Gemini 3.1 Pro \cite{team2023gemini} as a judge to provide a Gemini-Score (GS) based on real-world interior design standards. To evaluate the unique multi-story generation capability of Text2Villa, we introduce the Format Success Rate (FSR) to quantify the proportion of outputs that can be parsed into 3D buildings without errors. Additionally, we use the Valid Connection Rate (VCR) to verify whether stairs are correctly generated and floor slabs are properly connected between levels. At the micro-level, we report the Collision Rate (CR), Floating Rate (FR), and Out of Boundary Rate (OBR) to assess physical compliance. We specifically define the Containment Success Rate (CSR) to accurately measure fine-grained constraints such as books placed inside bookshelf compartments. For visual and semantic correctness, we employ the CLIP-Score (CS) to measure the semantic alignment between rendered multi-view images and texts. We also use GS to evaluate overall scene completeness and the rationality of interaction reserved areas, while counting the Number of Objects (NO) in the scene. All metrics represent the calculated average of three scenes for each architectural layout and room type.

\subsection{Quantitative Results}\label{subsec4.2}

\paragraph{Macro-Level} As shown in Table \ref{quantitative1}, our method comprehensively outperforms Holodeck and SceneFoundry in both RR and GS even in the single-story generation subtask. More importantly, when processing complex multi-story text descriptions, Text2Villa can still robustly output buildings with valid multi-story connectivity, showing comparable performance to MANSION. This demonstrates that our layout generator learns and follows real-world architectural logic.

\input{pic/userstudy}

\paragraph{Micro-Level} Table \ref{quantitative2} presents the fine-grained asset layout performance across nine different room types. LayoutVLM and SceneWeaver are highly prone to geometric collisions or floating issues when handling high-frequency physical interactions. In contrast, Text2Villa achieves physically compliant layouts with zero collisions and zero floating across all room types. When dealing with highly challenging object containment relationships, Holodeck and LayoutVLM fail to respond effectively and only manage simple top-surface support cases. SceneWeaver and MANSION show unstable performance despite having some processing capability. Benefiting from the A-PSSG and internal constraint solving, our method achieves an average CSR of 97.6\%. Furthermore, our scenes attain the best performance in semantic alignment and scene rationality even while generating a significantly larger number of assets than the baselines.

\paragraph{Human Preference} Table \ref{userstudy} reports the preference scores obtained from a blind test conducted by 20 human volunteers. Our method achieves an overwhelming advantage in both physical plausibility and semantic alignment, with an overall preference rate superior to all baselines. The implementation can be found in Appendix \ref{app:human}.

\subsection{Qualitative Results}\label{subsec4.3}

\input{pic/quantitative2}

\input{pic/qualitative2}

\paragraph{Macro-Level} Fig. \ref{qualitative0} provides a qualitative comparison of complete villa generation. Although MANSION supports multi-story generation, its grid-based orthogonal space partitioning prevents the synthesis of complex polygonal structures with arbitrary angles commonly found in building footprints. Furthermore, the floorplans generated via bubble diagrams occasionally result in implausible room connections and disproportionate area allocations. As illustrated in Fig. \ref{qualitative1}, Holodeck is limited to rectangular boundaries, generating fragmented room layouts that lack realistic architectural connectivity. SceneFoundry attempts to convert text into procedural parameters of Infinigen to drive layout generation. Nevertheless, its parameter mapping process suffers from deviations, which frequently result in imbalanced room area distribution and tend to produce unreasonable redundant windows. Conversely, Text2Villa successfully maps complex spatial descriptions such as open-plan kitchens into a globally connected layout with irregular polygonal boundaries. The generated 3D views exhibit extremely high villa-level architectural realism.

\paragraph{Micro-Level} Fig. \ref{qualitative2} provides a qualitative comparison of micro-level instantiation. Baselines exhibit obvious spatial and physical perception defects. While Holodeck produces physically reasonable global layouts, it cannot resolve fine-grained containment relationships and lacks visual realism in its rendering. MANSION can handle object containment to some extent by relying on a specific simulator and simple detection based on axis-aligned bounding boxes, but it lacks true 3D geometric awareness of internal cavities. Additionally, its generated scenes frequently exhibit semantic discrepancies from the text inputs. The differentiable optimization process of LayoutVLM fails to resolve all physical conflicts, leading to frequent intersections among furniture pieces. Its self-consistent decoding mechanism also struggles to correct orientation errors completely. SceneWeaver relies on agents that lack strict physical constraints, causing objects to scatter frequently on the floor away from supporting surfaces. This approach also bypasses collision issues through deletion operations, generating overly sparse scenes. In contrast, Text2Villa strictly enforces semantic orientation constraints such as placing nightstands on both sides of a double bed. It also achieves the precise placement of small objects inside display racks and storage cabinets, yielding highly organized, natural, and functional visual performances.

\input{pic/ablation}

\subsection{Ablation Study}\label{subsec4.4}

We verify the necessity of the key designs in Text2Villa, with quantitative results summarized in Table \ref{ablation}. 

\paragraph{Effectiveness of Fine-tuning} Regarding macro-level layout generation, removing the LoRA fine-tuning to rely solely on zero-shot LLMs causes a severe degradation in system performance. The FSR, which measures the successful parsing of output JSON sequences into 3D building meshes, plummets drastically from 98.2\% to 26.8\%. Furthermore, even when syntactically correct coordinate parameters are generated, the zero-shot model completely lacks spatial geometric reasoning, reflected by the drops in both RR and VCR. As qualitatively illustrated in Fig. \ref{wofine}, the model fails to translate textual instructions into irregular polygonal boundaries, generates unreachable enclosed rooms, and cannot establish valid multi-story staircase connections. This proves the necessity of constructing domain-specific datasets for autoregressive fine-tuning to learn implicit architectural rules.

\paragraph{Impact of A-PSSG} We revert the A-PSSG to a traditional pure semantic scene graph by removing physical affordance attributes and physical interaction edges. Under this configuration, the CR of the scenes surges to 36.3\%, the FR reaches 22.7\%, and the CSR completely drops to zero. This indicates that pure semantic relationships cannot guide precise geometric intersections in 3D space without explicit physical affordance abstraction. 

\input{pic/wofine}

\paragraph{Analysis of the Physics-Aware Closed-Loop Solver} We design three variants to evaluate the closed-loop instantiation mechanism. 1) W/o Closed-loop (Init-Only): Objects are coarsely placed only once based on the initial A-PSSG, without subsequent iterative optimization. The resulting scenes are filled with collisions and empty spaces, proving that feed-forward single pose estimation is highly unreliable in dense 3D spaces. 2) W/o Physics (MLLM-Only): The underlying physical check is removed, and pose updates are performed solely based on visual feedback from the MLLM. Because the MLLM lacks precise depth perception, it can move objects to semantically correct general areas but fails to handle fine-grained object interactions. 3) W/o MLLM (Physics-Only): The MLLM semantic misalignment penalty is discarded (by setting $\lambda_{sem}=0$), forcing the solver to search solely for the elimination of physical conflicts. Although the generated scenes are physically compliant, this setup triggers severe semantic confusion, such as televisions facing walls or beds blocking doorways. Reflected in the data, the GS evaluating semantic rationality reaches only 5.6.

%% file: pic/qualitative0.tex
\begin{figure*}[!htbp]
    \centering
    \includegraphics[width=\textwidth]{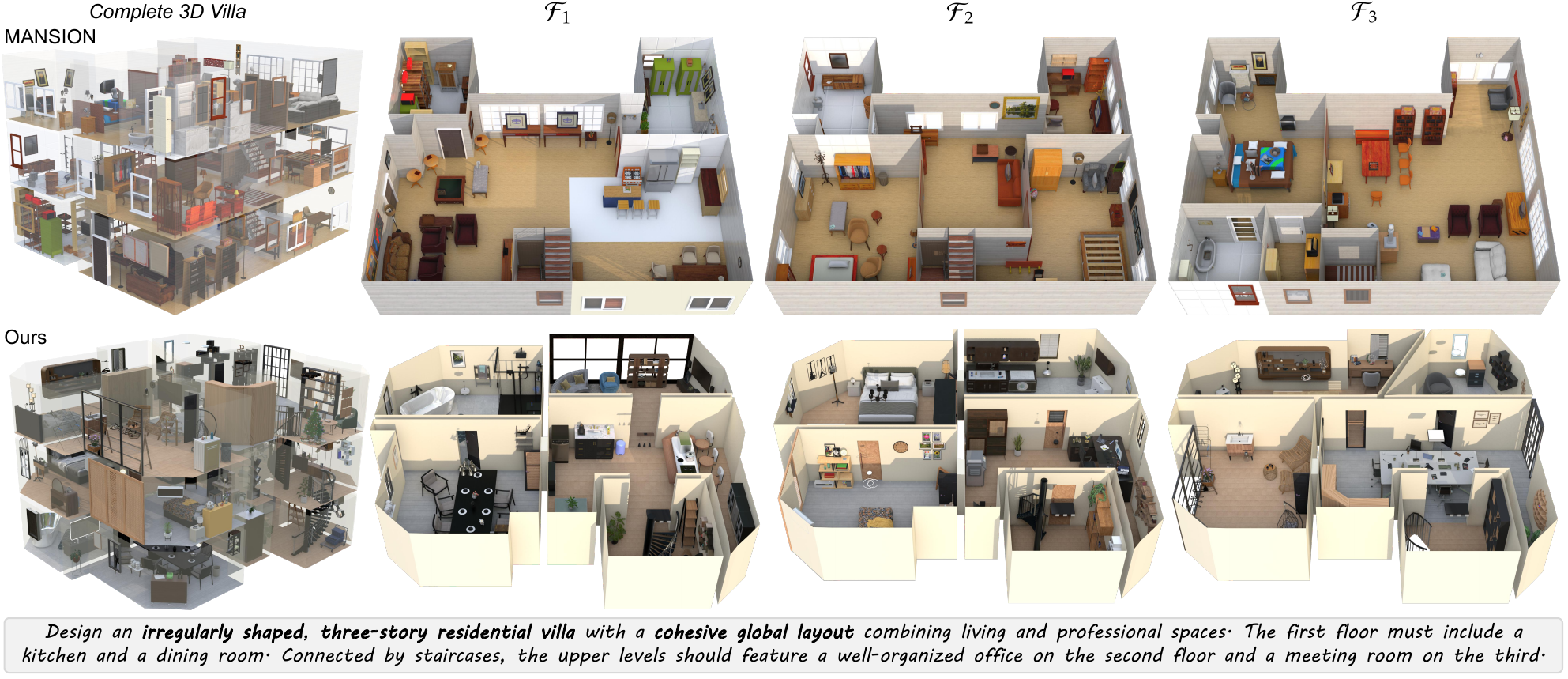}
    \vspace{-0.6cm}
    \caption{Qualitative comparison of complete villa layout generation. While MANSION also possesses multi-story generation capabilities, the building footprints produced by our method are not merely simple compositions and subtractions of rectangles. Moreover, Text2Villa can precisely generate the room categories specified in the text, equipped with plausible assets, and ultimately achieve photorealistic rendering results.}
    \label{qualitative0}
    \vspace{-0.4cm}
\end{figure*}

%% file: pic/quantitative1.tex
\begin{table}
    \centering
    \caption{Quantitative comparison of macro-level architectural layout generation. In the single-story subtask, our method achieves reachability of all rooms and a higher GS, indicating layouts that better align with real-world design standards. Furthermore, our approach demonstrates multi-story building generation capabilities comparable to MANSION.}
    \vspace{-0.3cm}
    \label{quantitative1}
    \resizebox{\columnwidth}{!}{%
    \begin{tabular}{l ccc c}
    \toprule
    \textbf{Method} & \textbf{RR (\%) $\uparrow$} & \textbf{GS $\uparrow$} & \textbf{FSR (\%) $\uparrow$} & \textbf{VCR (\%) $\uparrow$} \\
    \midrule
    Holodeck & 95.2 & 3.2 & - & - \\
    SceneFoundry & 97.7 & 7.3 & - & - \\
    \midrule
    MANSION (Single-story) & 100.0 & 6.4 & - & - \\
    MANSION (Multi-story) & 98.6 & 6.1 & - & 95.9 \\
    \midrule
    \textbf{Ours (Single-story)} & \textbf{100.0} & \textbf{7.6} & \textbf{99.1} & - \\
    \textbf{Ours (Multi-story)} & 98.9 & 7.4 & 98.2 & \textbf{97.3} \\
    \bottomrule
    \end{tabular}%
    }
    \vspace{-0.3cm}
\end{table}

%% file: pic/qualitative1.tex
\begin{figure*}[!htbp]
    \centering
    \includegraphics[width=\textwidth]{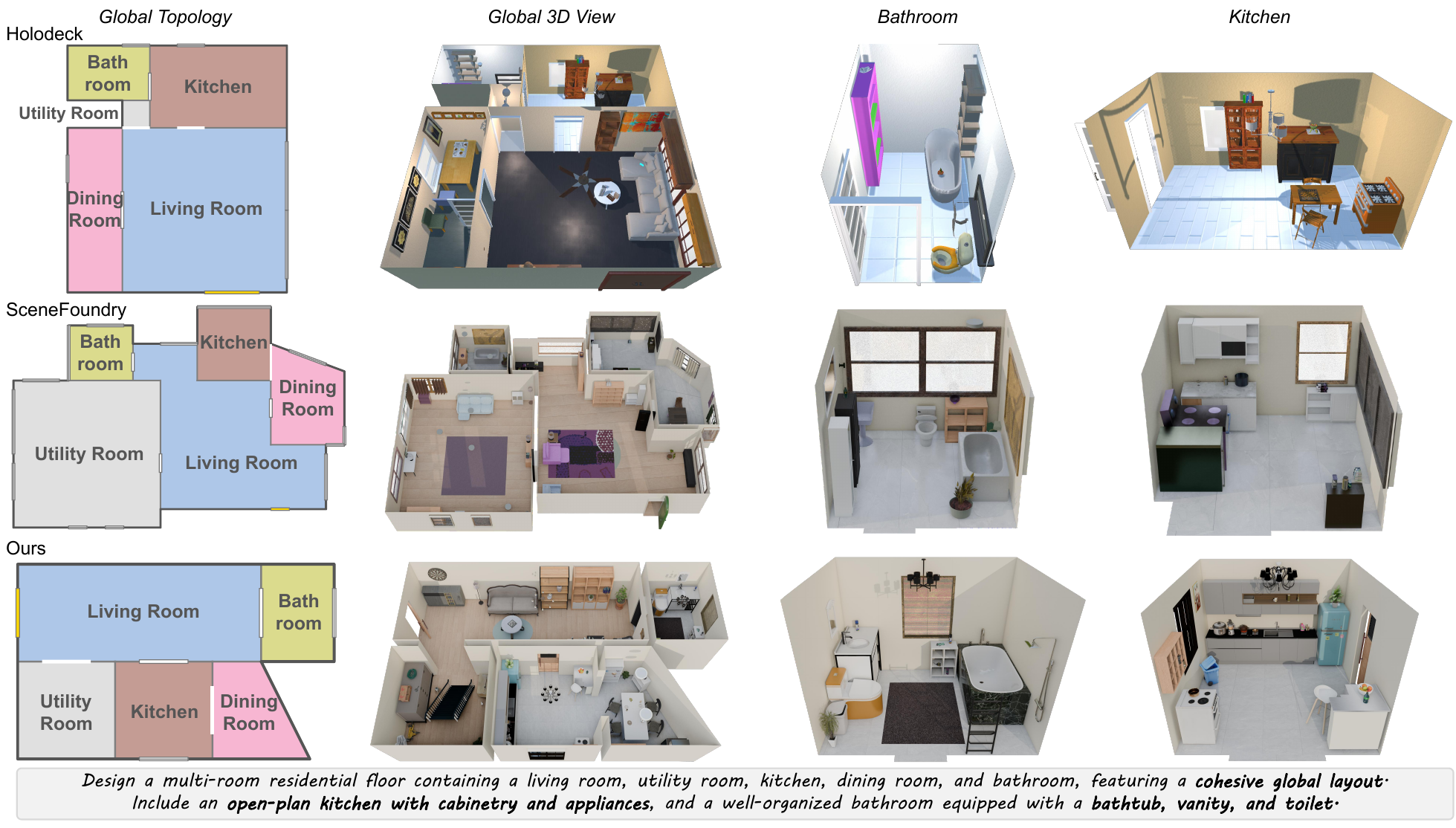}
    \vspace{-0.6cm}
    \caption{Qualitative comparison of macro-level architectural layout generation. Because baselines lack multi-story generation capabilities, we compare their multi-room outputs with the single-story degraded results of our approach. Our method robustly generates global room topologies compliant with real-world architectural design and accurately captures complex spatial directives from text prompts (e.g., an open-plan kitchen). The generated scenes exhibit high physical realism and semantic functionality in both macro-connectivity and micro-arrangements.}
    \label{qualitative1}
    \vspace{-0.4cm}
\end{figure*}

%% file: pic/userstudy.tex
\begin{table}[t]
    \centering
    \caption{Human Preference. Percentage of layout results generated by our method preferred by 20 human evaluators against baselines.}
    \vspace{-0.3cm}
    \label{userstudy}
    \resizebox{\columnwidth}{!}{%
    \begin{tabular}{l ccc}
    \toprule
    \textbf{Comparisons} & \textbf{Physical Plausibility} & \textbf{Semantic Alignment} & \textbf{Overall Preference} \\
    \midrule
    \textbf{Ours} vs. Holodeck & 67.7\% & 72.3\% & \textbf{79.1\%} \\
    \textbf{Ours} vs. MANSION & 62.7\% & 71.2\% & \textbf{68.8\%} \\
    \textbf{Ours} vs. LayoutVLM & 89.4\% & 88.7\% & \textbf{94.6\%} \\
    \textbf{Ours} vs. SceneWeaver & 84.0\% & 82.5\% & \textbf{89.8\%} \\
    \bottomrule
    \end{tabular}%
    }
    \vspace{-0.4cm}
\end{table}

%% file: pic/quantitative2.tex
\begin{table*}[htbp]
    \centering\small
    \caption{Quantitative comparison of micro-asset instantiation across 9 room types. Our method consistently outperforms existing open-vocabulary indoor scene generation methods, particularly in terms of physical plausibility and complex semantic constraints.}
    \vspace{-0.3cm}
    \setlength{\tabcolsep}{2pt} 
    \resizebox{\linewidth}{!}{%
    \begin{tabular}{l ccccccc ccccccc ccccccc ccccccc ccccccc}
    \toprule
    \multirow{2}{*}{\textbf{Method}} & \multicolumn{7}{c}{\textbf{Living Room}} & \multicolumn{7}{c}{\textbf{Bedroom}} & \multicolumn{7}{c}{\textbf{Bathroom}} & \multicolumn{7}{c}{\textbf{Kitchen}} & \multicolumn{7}{c}{\textbf{Dining Room}} \\
    \cmidrule(lr){2-8} \cmidrule(lr){9-15} \cmidrule(lr){16-22} \cmidrule(lr){23-29} \cmidrule(lr){30-36}
    & CR$\downarrow$ & FR$\downarrow$ & OBR$\downarrow$ & CSR$\uparrow$ & CS$\uparrow$ & GS$\uparrow$ & NO & CR$\downarrow$ & FR$\downarrow$ & OBR$\downarrow$ & CSR$\uparrow$ & CS$\uparrow$ & GS$\uparrow$ & NO & CR$\downarrow$ & FR$\downarrow$ & OBR$\downarrow$ & CSR$\uparrow$ & CS$\uparrow$ & GS$\uparrow$ & NO & CR$\downarrow$ & FR$\downarrow$ & OBR$\downarrow$ & CSR$\uparrow$ & CS$\uparrow$ & GS$\uparrow$ & NO & CR$\downarrow$ & FR$\downarrow$ & OBR$\downarrow$ & CSR$\uparrow$ & CS$\uparrow$ & GS$\uparrow$ & NO \\
    \midrule
    Holodeck    & 5.9 & 0.0 & 0.0 & 0.0 & 30.6 & 7.2 & 17.0 & 5.5 & 0.0 & 0.0 & 0.0 & 29.7 & 7.9 & 18.3 & 8.1 & 0.0 & 0.0 & 0.0 & 31.5 & 7.6 & 12.3 & 7.6 & 3.8 & 0.0 & 0.0 & 28.2 & 8.4 & \textbf{26.3} & 4.4 & 0.0 & 0.0 & 0.0 & 30.1 & 8.5 & 22.7 \\
    MANSION     & 0.0 & 0.6 & 0.0 & 89.1 & 31.4 & 7.1 & 24.5 & 0.0 & 0.0 & 0.0 & 92.5 & 33.1 & 8.9 & \textbf{24.3} & 0.0 & 0.8 & 0.0 & 92.6 & 32.9 & 8.4 & 13.8 & 0.0 & 0.0 & 0.0 & 92.1 & 29.3 & 8.1 & 21.9 & 0.0 & 0.0 & 0.0 & 87.2 & 28.5 & 7.9 & 25.6 \\
    LayoutVLM   & 21.5 & 10.7 & 0.0 & 0.0 & 25.3 & 2.7 & 9.3 & 14.6 & 0.0 & 0.0 & 0.0 & 28.2 & 7.3 & 13.7 & 10.3 & 20.6 & 0.0 & 0.0 & 26.4 & 5.8 & 9.7 & 11.6 & 5.9 & 0.0 & 0.0 & 26.9 & 7.2 & 17.3 & 6.3 & 25 & 6.3 & 0.0 & 27.6 & 3.1 & 16.0 \\
    SceneWeaver & 7.9 & 0.0 & 0.0 & 78.7 & 29.2 & 6.8 & 12.7 & 10.0 & 30.0 & 0.0 & 87.9 & 27.1 & 3.8 & 10.0 & 9.1 & 18.2 & 0.0 & 72.7 & 28.1 & 6.5 & 11.0 & 6.4 & 0.0 & 0.0 & 89.2 & 27.6 & 6.3 & 15.7 & 0.0 & 9.3 & 0.0 & 84.1 & 29.7 & 8.3 & 10.7 \\
    \textbf{Ours} & \textbf{0.0} & \textbf{0.0} & \textbf{0.0} & \textbf{96.3} & \textbf{34.5} & \textbf{8.9} & \textbf{27.0} & \textbf{0.0} & \textbf{0.0} & \textbf{0.0} & \textbf{100} & \textbf{33.4} & \textbf{9.1} & 23.7 & \textbf{0.0} & \textbf{0.0} & \textbf{0.0} & \textbf{95.2} & \textbf{34.2} & \textbf{9.0} & \textbf{14.7} & \textbf{0.0} & \textbf{0.0} & \textbf{0.0} & \textbf{95.8} & \textbf{33.6} & \textbf{8.8} & 24.0 & \textbf{0.0} & \textbf{0.0} & \textbf{0.0} & \textbf{96.4} & \textbf{32.7} & \textbf{8.9} & \textbf{36.3} \\
    \midrule
    \end{tabular}
    }
    \resizebox{\linewidth}{!}{%
    \begin{tabular}{l ccccccc ccccccc ccccccc ccccccc |ccccccc}
    \multirow{2}{*}{\textbf{Method}} & \multicolumn{7}{c}{\textbf{Childroom}} & \multicolumn{7}{c}{\textbf{Office}} & \multicolumn{7}{c}{\textbf{Meeting Room}} & \multicolumn{7}{c|}{\textbf{Utility Room}} & \multicolumn{7}{c}{\cellcolor{gray!20}\textbf{Average}} \\
    \cmidrule(lr){2-8} \cmidrule(lr){9-15} \cmidrule(lr){16-22} \cmidrule(lr){23-29} \cmidrule(lr){30-36}
    & CR$\downarrow$ & FR$\downarrow$ & OBR$\downarrow$ & CSR$\uparrow$ & CS$\uparrow$ & GS$\uparrow$ & NO & CR$\downarrow$ & FR$\downarrow$ & OBR$\downarrow$ & CSR$\uparrow$ & CS$\uparrow$ & GS$\uparrow$ & NO & CR$\downarrow$ & FR$\downarrow$ & OBR$\downarrow$ & CSR$\uparrow$ & CS$\uparrow$ & GS$\uparrow$ & NO & CR$\downarrow$ & FR$\downarrow$ & OBR$\downarrow$ & CSR$\uparrow$ & CS$\uparrow$ & GS$\uparrow$ & NO & CR$\downarrow$ & FR$\downarrow$ & OBR$\downarrow$ & CSR$\uparrow$ & CS$\uparrow$ & GS$\uparrow$ & NO \\
    \midrule
    Holodeck    & 4.9 & 4.9 & 0.0 & 0.0 & 31.7 & 7.3 & 20.3 & 3.0 & 0.0 & 0.0 & 0.0 & 31.2 & 8.2 & \textbf{33.0} & 4.8 & 0.0 & 0.0 & 0.0 & 29.6 & 8.1 & 20.7 & 0.0 & 0.0 & 0.0 & 0.0 & 22.5 & 3.4 & 3.3 & 4.9 & 1.0 & 0.0 & 0.0 & 29.5 & 7.4 & 19.3 \\
    MANSION     & 0.0 & 1.7 & 0.0 & 92.6 & 28.5 & 7.2 & 19.3 & 0.0 & 1.1 & 0.0 & 96.7 & 33.5 & 8.2 & 29.6 & 0.0 & 0.0 & 0.0 & 91.4 & 31.7 & 8.6 & \textbf{25.9} & 0.0 & 0.0 & 0.0 & 100 & 28.8 & 8.0 & 2.5 & 0.0 & 0.5 & 0.0 & 92.7 & 30.9 & 8.0 & 20.8 \\
    LayoutVLM   & 8.3 & 8.3 & 0.0 & 0.0 & 27.3 & 4.6 & 12.0 & 13.0 & 0.0 & 6.5 & 0.0 & 28.1 & 4.9 & 15.3 & 12.5 & 6.3 & 0.0 & 0.0 & 28.2 & 7.3 & 16.0 & 10.3 & 0.0 & 0.0 & 0.0 & 27.5 & 7.9 & 9.7 & 12.0 & 8.5 & 1.4 & 0.0 & 27.3 & 5.6 & 13.2 \\
    SceneWeaver & 13.6 & 0.0 & 0.0 & 88.4 & 26.5 & 4.2 & 14.7 & 7.7 & 23.0 & 0.0 & 92.3 & 27.1 & 5.8 & 13.0 & 7.9 & 0.0 & 0.0 & 86.6 & 26.8 & 7.7 & 12.7 & 0.0 & 20.0 & 0.0 & 100 & 28.3 & 7.8 & 5.0 & 7.0 & 11.2 & 0.0 & 86.7 & 27.8 & 6.4 & 11.7 \\
    \textbf{Ours} & \textbf{0.0} & \textbf{0.0} & \textbf{0.0} & \textbf{95.7} & \textbf{33.5} & \textbf{8.6} & \textbf{23.0} & \textbf{0.0} & \textbf{0.0} & \textbf{0.0} & \textbf{100} & \textbf{33.9} & \textbf{8.4} & 15.7 & \textbf{0.0} & \textbf{0.0} & \textbf{0.0} & \textbf{98.7} & \textbf{34.8} & \textbf{9.2} & 22.3 & \textbf{0.0} & \textbf{0.0} & \textbf{0.0} & \textbf{100} & \textbf{29.6} & \textbf{8.2} & \textbf{12.0} & \textbf{0.0} & \textbf{0.0} & \textbf{0.0} & \textbf{97.6} & \textbf{33.4} & \textbf{8.8} & \textbf{22.0} \\
    \bottomrule
    \end{tabular}
    }
    \label{quantitative2}
    \vspace{-0.3cm}
\end{table*}

%% file: pic/qualitative2.tex
\begin{figure*}
    \centering
    \includegraphics[width=\textwidth]{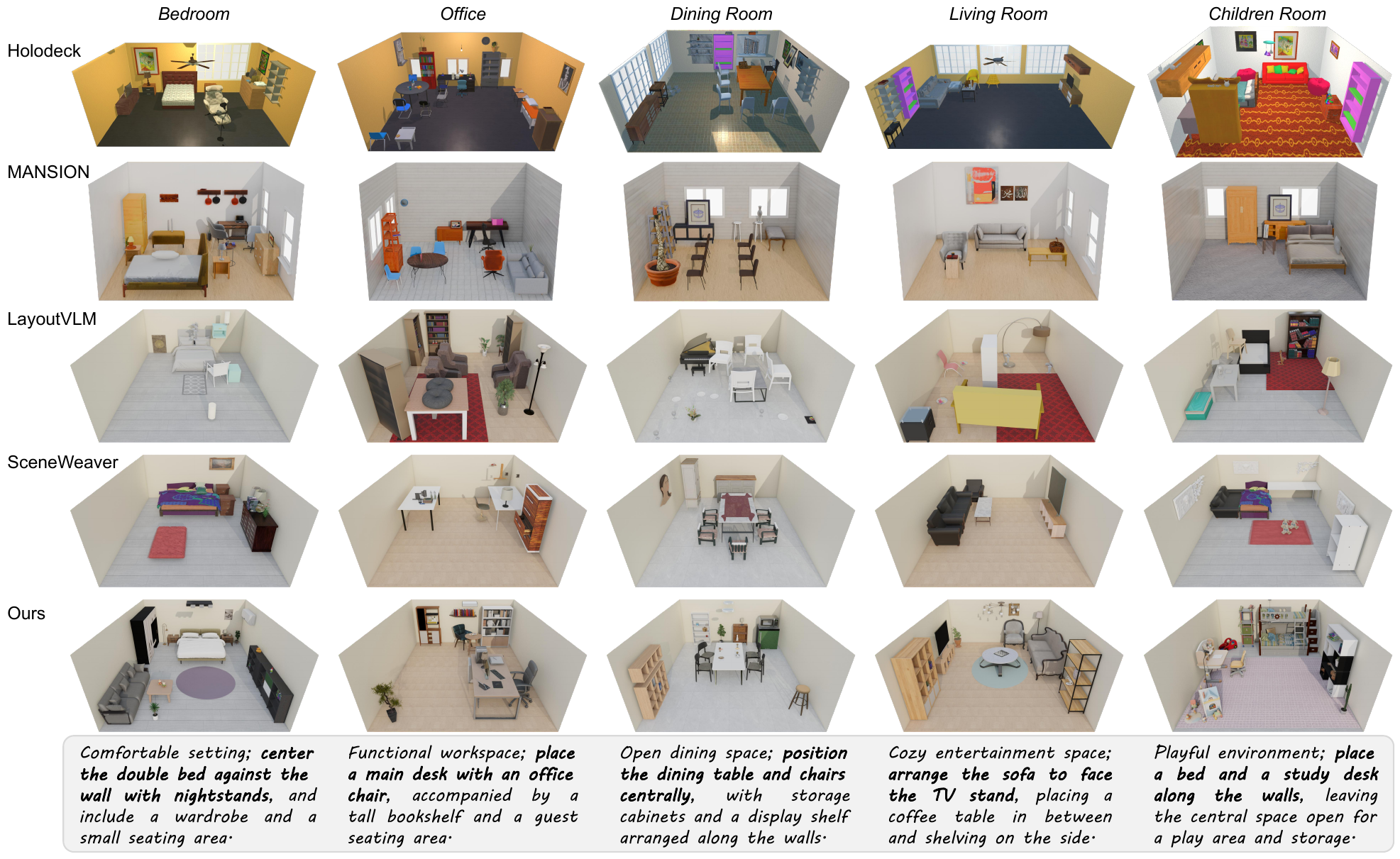}
    \vspace{-0.6cm}
    \caption{Qualitative comparison of micro-level asset instantiation across different room types. We compare against baselines to evaluate the capability of precise 3D asset arrangement based on detailed natural language instructions. Our method generates scene layouts that closely adhere to prompts while satisfying fine-grained physical and semantic dual constraints.}
    \vspace{-0.4cm}
    \label{qualitative2}
\end{figure*}

%% file: pic/ablation.tex
\begin{table}
    \centering
    \caption{Quantitative Ablation Study of key designs in Text2Villa.}
    \vspace{-0.3cm}
    \label{ablation}
    \resizebox{\columnwidth}{!}{%
    \begin{tabular}{l ccc ccccccc}
    \toprule
    \multirow{2}{*}{\textbf{Model Variants}} & \multicolumn{3}{c}{\textbf{Macro-Topology}} & \multicolumn{7}{c}{\textbf{Micro-Instantiation}} \\
    \cmidrule(lr){2-4} \cmidrule(lr){5-11}
    & RR$\uparrow$ & FSR$\uparrow$ & VCR$\uparrow$ 
    & CR$\downarrow$ & FR$\downarrow$ & OBR$\downarrow$ 
    & CSR$\uparrow$ & CS$\uparrow$ & GS$\uparrow$ & NO \\
    \midrule
    W/o LLM Finetune & 76.9 & 26.8 & 80.2 & - & - & - & - & - & - & - \\
    \midrule
    W/o A-PSSG & - & - & - & 36.3 & 22.7 & 0.0 & 0.0 & 31.6 & 8.1 & 22.0 \\
    \midrule
    W/o Closed-loop & - & - & - & 27.2 & 22.7 & 0.0 & 0.0 & 29.5 & 3.7 & 8.0 \\
    W/o Physics & - & - & - & 31.8 & 27.2 & 0.0 & 0.0 & 30.3 & 7.4 & 22.0 \\
    W/o MLLM & - & - & - & 9.1 & 4.5 & 0.0 & 45.5 & 28.7 & 5.6 & 8.0 \\
    \midrule
    \textbf{Ours} 
    & \textbf{98.9} & \textbf{98.2} & \textbf{97.3} 
    & \textbf{0.0} & \textbf{0.0} & \textbf{0.0} 
    & \textbf{97.6} & \textbf{33.4} & \textbf{8.8} & \textbf{22.0} \\
    \bottomrule
    \end{tabular}
    }
    \vspace{-0.3cm}
\end{table}

%% file: pic/wofine.tex
\begin{figure}[t]
    \centering
    \includegraphics[width=\columnwidth]{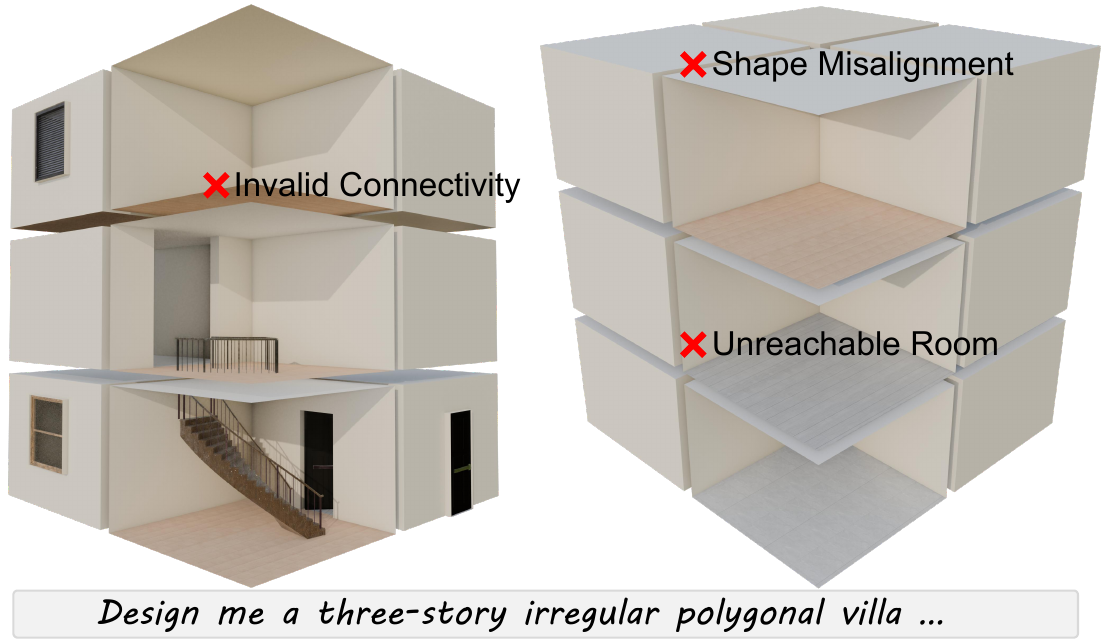}
    \vspace{-0.6cm}
    \caption{Qualitative ablation on macro-level architectural layout generation.}
    \vspace{-0.4cm}
    \label{wofine}
\end{figure}

%% file: sec/5_discussion.tex
\section{Discussion}

\paragraph{Scene Energy Convergence Analysis}

Text2Villa experiences performance degradation when handling extremely dense asset instantiation. Concretely, when the input instruction requires placing an excessive number of objects on a supporting surface with limited area (e.g., a small coffee table), the solver encounters a conflict between the  continuous physical space capacity and the discrete atomic actions. Fig. \ref{failure} illustrates a failure case. During the closed-loop optimization, as the solver iteratively invokes \texttt{add\_entity} to insert new objects, the available tabletop space gradually diminishes. New objects inevitably cause volumetric penetrations with already placed items (increasing $E_{col}$) or exceed the support boundaries (increasing $E_{sup}$). Although the solver attempts to find valid poses via \texttt{update\_pose}, the hard constraints of the physical space cause the support and collision penalty terms to oscillate repeatedly across multiple iterations, preventing both from converging to zero simultaneously. To break this deadlock and force a reduction in the total scene energy $E(S_t)$, the solver ultimately triggers \texttt{remove\_entity} to delete some assets and free up space. Consequently, the final generated scene is valid and collision-free, yet exhibits a semantic discrepancy regarding the exact quantity specified in the user prompt. This reveals the limitations of the current compromise through penalties and highlights an important direction for our future exploration in complex scene optimization.

\input{pic/failure}

\paragraph{Future Work} Text2Villa achieves remarkable progress in generating multi-story buildings and opens up promising avenues for future exploration. Regarding asset acquisition, although we currently rely on finite databases, achieving fully open-vocabulary expansion can be readily addressed by integrating emerging native 3D generation models on the fly. Furthermore, real-world indoor scenes encompass a myriad of deformable (e.g., cloth) and articulated objects (e.g., cabinet doors). Although our proposed A-PSSG primarily focuses on the most common support, cavity containment, and functional orientation in static rigid body scenes, its graph structure is inherently extensible. Mapping kinematic parameters into node attributes and integrating them into the closed-loop solver remains a valuable open problem. From a computational perspective, the multiple rounds of heuristic search executed by the solver incur substantial inference latency (see Appendix \ref{app:runtime} for runtime analysis), particularly when resolving complex interactions among dozens of microscopic assets. We plan to implement parallel optimization across rooms and explore knowledge distillation to train a lightweight, dedicated 3D spatial reasoning network, thereby drastically accelerating the convergence of scene instantiation.

\paragraph{Application}

As illustrated in Fig. \ref{application}, Text2Villa can generate villa-scale 3D environments from text, demonstrating significant potential for various downstream applications. For game development, Text2Villa lowers the barrier to user-generated content. The generated architectural assets and furniture are represented as explicit meshes that seamlessly integrate into mainstream game engines such as Unity and Unreal Engine for rapid prototyping. This integration effectively reduces the manual modeling costs typically required to build large-scale virtual communities. Driven by the dual physical and semantic constraints of A-PSSG alongside a physics-aware closed-loop solver, Text2Villa produces highly realistic scenes with zero collisions and accurate affordances. This capability establishes a reliable and interactive real-to-sim workflow, providing critical support for training embodied agents in complex multi-room tasks.

\input{pic/application}

%% file: pic/failure.tex
\begin{figure}[t]
    \centering
    \includegraphics[width=\columnwidth]{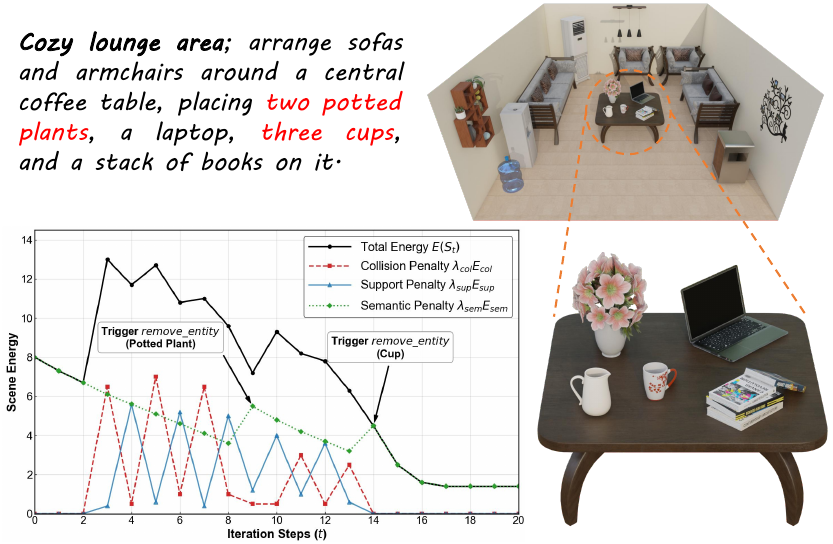}
    \vspace{-0.6cm}
    \caption{A failure case.}
    \vspace{-0.4cm}
    \label{failure}
\end{figure}

%% file: pic/application.tex
\begin{figure}[t]
    \centering
    \includegraphics[width=\columnwidth]{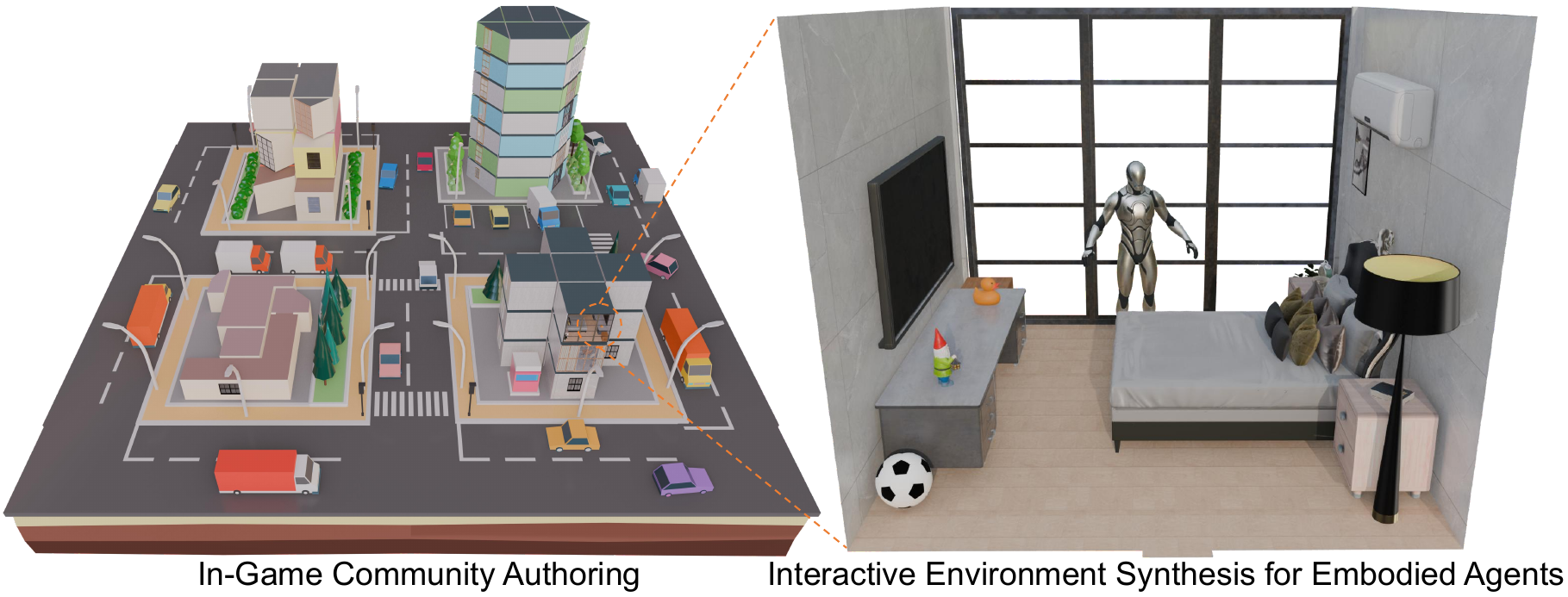}
    \vspace{-0.6cm}
    \caption{Downstream applications of Text2Villa. The four multi-story buildings with distinct shapes in the left figure are generated by our method.}
    \vspace{-0.4cm}
    \label{application}
\end{figure}

%% file: sec/6_conclusion.tex
\section{Conclusion}

In this paper, we present Text2Villa, a text-driven multi-story 3D indoor scene generation framework. To bridge the gap between highly abstract natural language and strict physical reality, we bypass the limitations of forward generation paradigms and single-room assumptions by proposing a hierarchical generation pipeline. At the macro level, we fine-tune an autoregressive layout generator that parses unstructured text to generate architectural foundations containing polygonal boundaries and multi-story connectivity. At the micro level, we expand the macro nodes and propose A-PSSG, which extends traditional pure semantic relations into physical interaction graphs encompassing geometric support and containment cavity constraints. Building upon this representation, we design a closed-loop iterative solver rooted in the analysis-by-synthesis paradigm. By integrating semantic feedback from MLLM with the hard constraints of an underlying physics engine, this solver adopts a dynamic observation-evaluation-modification mechanism, effectively resolving common issues such as collision, floating, and semantic error during fine-grained asset instantiation within a continuous parameter space. Experiments demonstrate that Text2Villa achieves high-quality generation from text to villa-scale architectures and provides a robust algorithmic framework for the physically compliant placement of assets. Text2Villa offers a reliable pathway for generating coherent and physically usable 3D scenes for applications in virtual reality, digital content generation, and embodied artificial intelligence.

%% file: sec/7_appendix.tex
\clearpage
\appendix

\twocolumn[
  \begin{center}
    {\Large \bf Text2Villa: Hierarchical Generation of 3D Indoor Environments with Physics-Aware Analysis-by-Synthesis \par} 
    \vspace{1em} 
    {\Large \bf Supplementary Material \par} 
    \vspace{1.5em} 
  \end{center}
]

\section{Dataset Construction Details}\label{app:dataset}
As mentioned in Subsec. \ref{subsec3.1}, we construct a fine-tuning dataset containing 1,000 text-JSON pairs to train the autoregressive layout generator. To this end, we develop a procedural spatial partitioner that synthesizes geometrically precise building foundations and floor plans.

\paragraph{Procedural Spatial Partitioner} 
The spatial partitioner relies on 2D Boolean operations and the binary space partitioning (BSP) algorithm. The process operates as follows:
\begin{enumerate}
    \item Exterior Contour Initialization: We start with a rectangular bounding box. To create diverse architectural footprints (e.g., L-shape, U-shape, T-shape, H-shape, and irregular polygons), we subtract randomized rectangular chunks from the corners or edges using Boolean difference operations via the \texttt{shapely} library.
    \item Internal BSP Partitioning: The remaining polygon on each floor is recursively subdivided using a BSP tree. Splitting hyperplanes are randomly selected along the $X$ or $Y$ axes, terminating when the target number of rooms is reached or the area falls below a usability threshold.
    \item Layout Connectivity and 1D Elements: Based on the partitioned 2D rooms, we extract shared line segments between adjacent rooms as interior walls, and segments without neighbors as exterior walls. \texttt{Doors} and \texttt{opens} (represented as 1D \texttt{LineString}) are placed on interior walls to establish accessibility based on functional priors. \texttt{Windows} are sampled on exterior walls, and a main \texttt{entrance} is assigned to an exterior wall on the ground floor.
    \item Vertical Alignment: For multi-story buildings, a fixed $4\times4$ staircase is rigidly anchored in the identical room across all floors, thereby guaranteeing valid vertical multi-story connectivity.
\end{enumerate}

\paragraph{Human-in-the-loop Annotation}
The raw outputs generated by the spatial partitioner are mathematically precise but lack semantic information. We adopt a human-in-the-loop approach to assign specific room categories to the layout, such as labeling large polygonal areas as living rooms, and to filter out unreasonable configurations. To enhance the generalization capability of the model to diverse user instructions, we annotate each layout with either coarse-grained or fine-grained natural language descriptions.

\paragraph{Dataset Statistics and Setup}
The dataset contains 1,000 samples in total, with a ratio of 2:4:4 for single-story, two-story, and three-story buildings. Table \ref{textjsonpair} presents an example of a typical text-JSON data pair. Although the procedural generation theoretically supports buildings with unlimited floors, we restrict the maximum number of floors to three to balance the computational overhead of micro-asset instantiation. Each floor has a standard height of 2.8 meters and contains an average of five rooms. While these data do not yet cover all architectural designs, they adequately encompass the common building foundations (such as L-shaped, U-shaped, and irregular polygons) in the field of floor plan and multi-room generation. Our goal is to let the layout generator learn the distribution of the procedural partitioner, including structural rules for multi-room, multi-story, arbitrary polygonal boundaries, and vertical connectivity, thereby allowing it to generalize to unseen text prompts (e.g., star-shaped, hexagon). The system prompts used to fine-tune the autoregressive layout generator are provided in Table \ref{prompt1}. Fig. \ref{layout} visualizes diverse 3D building foundations generated by Text2Villa, which are obtained by processing the JSON data from the layout generator through its geometric parsing module built based on Infinigen \cite{raistrick2024infinigen}. The generated layouts overcome the limitations of traditional single-room bounding boxes and smoothly scale from single-story apartments to multi-story villas.

\input{pic/pair}

\input{pic/prompt1}

\input{pic/layout}

\input{pic/human}

\section{Custom Metrics}\label{app:metrics}

Since existing indoor generation metrics (e.g., IoU or FID) struggle to effectively evaluate complex architectural connectivity and micro-level physical interactions, we define and apply the following custom metrics.

RR is computed based on a graph traversal algorithm to detect the presence of unreachable enclosed rooms on each floor. Given the spatial layout graph $(\mathcal{R}, \mathcal{E})$ for floor $\mathcal{F}_i$, we initiate a breadth-first search from the \texttt{Utility Room} node to obtain the set of reachable nodes $\mathcal{R}_{reach}$ within its connected subgraph:
\begin{equation}
    \text{RR} = \frac{|\mathcal{R}_{reach}|}{|\mathcal{R}|}
\end{equation}
GS follows the MLLM scoring paradigm adopted in recent state-of-the-art works \cite{yang2025sceneweaver, xia2026sage} to evaluate the overall richness and visual plausibility of the rendered scenes. FSR calculates the proportion of JSON sequences output by the layout generator that are free of syntax errors, successfully parsed, and mapped into valid 3D building footprints. VCR verifies the vertical connectivity within multi-story building models. For any two adjacent floors $\mathcal{F}_i$ and $\mathcal{F}_{i+1}$, let $P_{staircase}^{(i)}$ denote the 2D contour polygon of the staircase on floor $i$, and $P_{hole}^{(i+1)}$ be the hole polygon on the floor slab of floor $i+1$. We determine connectivity validity by computing the intersection area of their horizontal projections:
\begin{equation}
\resizebox{0.9\linewidth}{!}{$ \displaystyle
    \text{VCR} = \frac{1}{N-1} \sum_{i=1}^{N-1} \mathbb{I} \Big( \text{Area}(P_{staircase}^{(i)} \cap P_{hole}^{(i+1)}) > 0 \Big)
$}
\end{equation}
CR, FR, and OBR are adapted and improved from SceneEval \cite{tam2026sceneeval}. Specifically, CR quantifies the volumetric intersection between independent meshes in a scene. FR judges whether the distance between the bottom surface of an object and its nearest supporting surface exceeds a threshold $\tau_{dist}$ (set to 0.01m). OBR checks whether the bounding box of a placed object extends entirely beyond the projected boundary of its supporting surface. Their specific formulations are given by the energy terms $E_{col}$ and $E_{sup}$ in Subsec. \ref{subsec3.4}. CSR precisely evaluates cavity containment relationships. It not only verifies whether the 3D bounding box $B_{child}$ of the child object $O_i$ is contained within the internal cavity $\mathcal{C}_k$ of the parent object $O_j$, but also requires that $O_i$ is stably placed on the cavity bottom surface $\mathcal{S}_{\mathcal{C}_k}$ without any solid mesh penetration:
\begin{equation}
\begin{split}
\text{CSR} &= \frac{1}{|\mathcal{E}_{phys:inside}|} \sum_{(i,j) \in \mathcal{E}_{phys:inside}} \\
&\quad \mathbb{I} \Big( (B_{child} \subseteq \mathcal{C}_k) \land (\text{Vol}(O_i \cap O_j) = 0) \\
&\quad \land (\| p_{bottom}^i - p_{proj}^i \|_2 \le \tau_{dist}) \Big)
\end{split}
\end{equation}

\section{Human Preference}\label{app:human}

We conduct a user study to comprehensively evaluate the quality of micro-level scene instantiation from the perspective of human perception. We invite twenty volunteers to evaluate ten distinct indoor scenes through a custom web-based interface shown in Fig. \ref{human}. For each case, the interface presents the input text description at the top and displays the 3D scenes generated by four baselines and Text2Villa side-by-side at the bottom. To eliminate potential bias caused by the presentation order, these five rendering results are randomized for each trial. Participants are asked to carefully observe the scenes and select the optimal result according to three criteria, namely physical plausibility, semantic consistency, and overall preference. The interface provides clear definitions for each criterion to ensure that volunteers apply consistent evaluation standards. We ultimately collect 600 independent and valid votes. The detailed comparisons are summarized in Table \ref{userstudy}.

\section{Runtime Analysis}\label{app:runtime}

Generating fully instantiated, physically and semantically plausible 3D scenes is a complex combinatorial optimization problem. As detailed in Subsec. \ref{subsec3.4}, Text2Villa employs an analysis-by-synthesis paradigm driven by a heuristic closed-loop solver. While this design eliminates physical conflicts like collision and floating prevalent in feed-forward methods, it introduces higher inference latency.

\paragraph{Time-Quality Trade-off} Experiments show that the energy function $E(S_t)$ for micro-level asset instantiation typically converges within 20 iterations. Under the predefined safety threshold $E(S_t) \leq 1.0$, physical conflicts are completely eliminated, leaving only slight and acceptable semantic deviations. The average instantiation time per room is approximately 15 minutes, resulting in a total time of about 225 minutes for a standard three-story villa comprising approximately 15 rooms. We report the single-room generation time of each method in Table \ref{time}. The major computational bottleneck arises at each optimization step $t$: the solver must query the MLLM via API and wait for its response to compute $E_{sem}(S_t)$ and determine the next action. This frequent network communication overhead significantly prolongs the total running time. We consider this a necessary and worthwhile trade-off: exchanging inference time for absolute physical correctness, effectively elevating the usability of generated scenes from visually plausible to physically interactive for downstream embodied AI tasks.

\input{pic/time}

\paragraph{Mitigation and Scalability}
The computational cost primarily stems from continuous MLLM API calls, which cost approximately \$0.80 per room, and the massive search space of 6-DoF poses. We provide the following strategies to enhance runtime flexibility:
\begin{itemize}
    \item Search Space Reduction: Layout generation for small rooms takes significantly less time than for open areas, because spatial boundaries heavily restrict the number of valid heuristic actions.
    \item Hyperparameter tuning: Users can accelerate the generation process by reducing the maximum number of iterations or increasing the energy safety threshold $E(S_t)$. Although this adjustment might affect the results, the underlying physics engine ensures that any degradation in layout quality remains marginal and controllable.
\end{itemize}

\input{pic/prompt2}

%% file: pic/pair.tex
\begin{figure*}[!ht]\centering
\captionof{table}{A text-JSON pair in the dataset.}
\begin{minipage}{\textwidth}\vspace{0mm}    \centering
\begin{tcolorbox}[llmprompt]
\small

\textbf{Text:} I want to build a three-story, irregularly shaped villa. The house needs a living room, a kitchen, a dining room, a master bedroom, a children's room, a meeting room, an office, two bathrooms, three utility rooms, and two balconies. For the layout, I hope the kitchen on the first floor can be openly connected to the dining room, and the living room can directly access the kitchen. Both the second-floor bedroom and the third-floor children's room should directly connect to balconies with panoramic windows. In addition, please include a TV in the meeting room and place a double bed in the children's room.

\vspace{2mm}

\textbf{JSON:}\vspace{-2mm}
{\scriptsize
\begin{verbatim}
{
  "rooms": {
      "utility-room_0/0": {"shape": "shapely.box(0, 0, 4, 4)"},
      "kitchen_0/0": {"shape": "shapely.box(4, 0, 8, 4)"},
      "dining-room_0/0": {"shape": "shapely.Polygon([(8, 0), (12, 0), (10, 4), (8, 4)])"},
      "living-room_0/0": {"shape": "shapely.box(0, 8, 20, 16)"},
      "bathroom_0/0": {"shape": "shapely.Polygon([(10, 4), (13, 4), (13, 8), (10, 8)])"},
      "utility-room_1/0": {"shape": "shapely.box(0, 0, 4, 4)"},
      "bedroom_1/0": {"shape": "shapely.box(4, 0, 8, 4)"},
      "meeting-room_1/0": {"shape": "shapely.box(0, 4, 8, 8)"},
      "office_1/0": {"shape": "shapely.Polygon([(8, 4), (11, 4), (11, 8), (8, 8)])"},
      "balcony_1/0": {"shape": "shapely.Polygon([(8, 0), (10, 2), (8, 4)])"},
      "utility-room_2/0": {"shape": "shapely.box(0, 0, 4, 4)"},
      "children-room_2/0": {"shape": "shapely.box(0, 4, 8, 8)"},
      "bathroom_2/0": {"shape": "shapely.box(4, 0, 8, 4)"},
      "balcony_2/0": {"shape": "shapely.Polygon([(8, 4), (10, 6), (8, 8)])"}
  },
  "entrance": {
      "main_door": {"shape": "shapely.LineString([(0, 5), (0, 7)])", "level": 0}
  },
  "doors": {
      "living0_kitchen0": {"shape": "shapely.LineString([(5, 4), (7, 4)])", "level": 0},
      "living0_bathroom0": {"shape": "shapely.LineString([(10, 5), (10, 7)])", "level": 0},
      "meeting1_bedroom1": {"shape": "shapely.LineString([(5, 4), (7, 4)])", "level": 1},
      "meeting1_office1": {"shape": "shapely.LineString([(8, 5), (8, 7)])", "level": 1},
      "bedroom1_balcony1": {"shape": "shapely.LineString([(8, 1), (8, 3)])", "level": 1},
      "children2_bathroom2": {"shape": "shapely.LineString([(5, 4), (7, 4)])", "level": 2},
      "children2_balcony2": {"shape": "shapely.LineString([(8, 5), (8, 7)])", "level": 2}
  },
  "opens": {
      "utility0_living0": {"shape": "shapely.LineString([(1, 4), (3, 4)])", "level": 0},
      "kitchen0_dining0": {"shape": "shapely.LineString([(8, 1), (8, 3)])", "level": 0},
      "utility1_meeting1": {"shape": "shapely.LineString([(1, 4), (3, 4)])", "level": 1},
      "utility2_children2": {"shape": "shapely.LineString([(1, 4), (3, 4)])", "level": 2}
  },
  "windows": {
      "kitchen0_window": {"shape": "shapely.LineString([(5, 0), (7, 0)])", "is_panoramic": 0, "level": 0},
      "bathroom0_window": {"shape": "shapely.LineString([(13, 5), (13, 7)])", "is_panoramic": 0, "level": 0},
      "office1_window": {"shape": "shapely.LineString([(11, 6), (11, 7)])", "is_panoramic": 0, "level": 1},
      "balcony1_pano": {"shape": "shapely.LineString([(8, 0), (10, 2)])", "is_panoramic": 1, "level": 1},
      "bathroom2_window": {"shape": "shapely.LineString([(5, 0), (7, 0)])", "is_panoramic": 0, "level": 2},
      "balcony2_pano": {"shape": "shapely.LineString([(8, 4), (10, 6)])", "is_panoramic": 1, "level": 2}
  },
  "interiors": {
      "wall_utility0_kitchen0": {"shape": "shapely.LineString([(4, 0), (4, 4)])", "level": 0},
      "wall_utility0_living0": {"shape": "shapely.LineString([(0, 4), (4, 4)])", "level": 0},
      "wall_living0_kitchen0": {"shape": "shapely.LineString([(4, 4), (8, 4)])", "level": 0},
      "wall_living0_dining0": {"shape": "shapely.LineString([(8, 4), (10, 4)])", "level": 0},
      "wall_kitchen0_dining0": {"shape": "shapely.LineString([(8, 0), (8, 4)])", "level": 0},
      ...
  }
}
\end{verbatim}
}

\end{tcolorbox}
\end{minipage}
\label{textjsonpair}
\end{figure*}

%% file: pic/prompt1.tex
\begin{figure*}[!ht]\centering
\captionof{table}{System prompt for fine-tuning the autoregressive layout generator.}
\begin{minipage}{\textwidth}\vspace{0mm}    \centering
\begin{tcolorbox}[llmprompt]
\small

You are an experienced architect and algorithm developer specializing in procedural floor plan generation. Your task is to generate 2D floor plans based on user requirements and strictly output them in JSON format parsed by Python's \texttt{shapely} library.

\textbf{Coordinate System \& Geometric Rules:}
\begin{itemize}
    \item Work in a 2D Cartesian coordinate system ($X$, $Y$ in meters).
    \item Use \texttt{shapely.box(min\_x, min\_y, max\_x, max\_y)} or \texttt{shapely.Polygon([(x1,y1), ...])} to define areas. The JSON must contain at least one polygonal room defined by \texttt{Polygon}.
    \item Use \texttt{shapely.LineString([(x1,y1), (x2,y2)])} for 1D elements (\texttt{entrance}, \texttt{doors}, \texttt{opens}, \texttt{windows}, \texttt{interiors}).
    \item Rooms must fit together perfectly with absolutely no gaps.
\end{itemize}

\textbf{JSON Structure \& Topology:}
Output a JSON object with 6 keys: \texttt{rooms}, \texttt{entrance}, \texttt{doors}, \texttt{opens}, \texttt{windows}, \texttt{interiors}.
\begin{itemize}
    \item \textbf{Naming}: Format must be \texttt{roomname\_x/y} (e.g., \texttt{kitchen\_1/1} is the 2nd kitchen on the 2nd floor).
    \item \textbf{Level Identifiers}: 1D elements must contain a \texttt{level} key (integer $x$ for the $(x+1)$-th floor). \texttt{entrance} must be level $0$. Missing level means shared across all floors.
    \item \textbf{Topology Constraints}: Line segments must lie strictly on the shared boundaries of the connected \texttt{rooms}. All rooms must be fully accessible (no enclosed/isolated spaces).
\end{itemize}

\textbf{Architectural Specifications:}
\begin{enumerate}
    \item \textbf{Layout}: Max 3 floors, max 6 rooms. For multi-story buildings, the external contours may vary, but the upper floor size must be smaller than the lower one; and the internal layout should be diversified.
    \item \textbf{Balcony \& Windows}: Balconies must have panoramic \texttt{windows} (\texttt{"is\_panoramic": 1}) and connect via \texttt{doors}. Kitchens and bathrooms must have \texttt{windows}.
    \item \textbf{Connections}: Kitchens connect to dining rooms. Living rooms connect to kitchens or dining rooms. Nested rooms must use \texttt{doors}.
    \item \textbf{Staircase}: Multi-story buildings must be equipped with staircases, which are fixed in the utility rooms with a uniform size of $4 \times 4$. Connect via \texttt{opens}. Avoid placing in the exact center.
    \item \textbf{Sizing}: Maintain reasonable sizes (larger living-rooms/bedrooms, smaller kitchens/bathrooms).
\end{enumerate}

\textbf{Task:} Design the layout and provide precise coordinates. Ensure topological correctness and accessibility.
Generate \textbf{one} JSON room layout templates with diverse layouts. Return ONLY the JSON format as shown below, without Markdown code blocks:
\vspace{-2mm}
{\scriptsize
\begin{verbatim}  
{
  "rooms": {
      "living-room_0/0": {"shape": "shapely.box(0, 0, 6, 5)"},
      "bedroom_0/0": {"shape": "shapely.Polygon([(6, 0), (9, 0), (9,3), (11, 3), (11, 5), (6, 5)])"},
      "staircase-room_0/0": {"shape": "shapely.box(0, 5, 4, 9)"}},
  "entrance": {"main_door": {"shape": "shapely.LineString([(2, 0), (4, 0)])", "level": 0}},
  "doors": {"living0_bedroom0": {"shape": "shapely.LineString([(6, 1), (6, 3)])", "level": 0}},
  "opens": {"utility0_living0": {"shape": "shapely.LineString([(1, 5), (3, 5)])", "level": 0}},
  "windows": {"bedroom0_window": {"shape": "shapely.LineString([(7, 5), (9, 5)])", "is_panoramic": 0, "level": 0}},
  "interiors": {
      "wall_utility0_living0": {"shape": "shapely.LineString([(0, 5), (4, 5)])", "level": 0},
      "wall_living0_bedroom0": {"shape": "shapely.LineString([(6, 0), (6, 5)])", "level": 0}}
}
\end{verbatim}
}

\end{tcolorbox}
\end{minipage}
\label{prompt1}
\end{figure*}

%% file: pic/layout.tex
\begin{figure*}
    \centering
    \includegraphics[width=\textwidth]{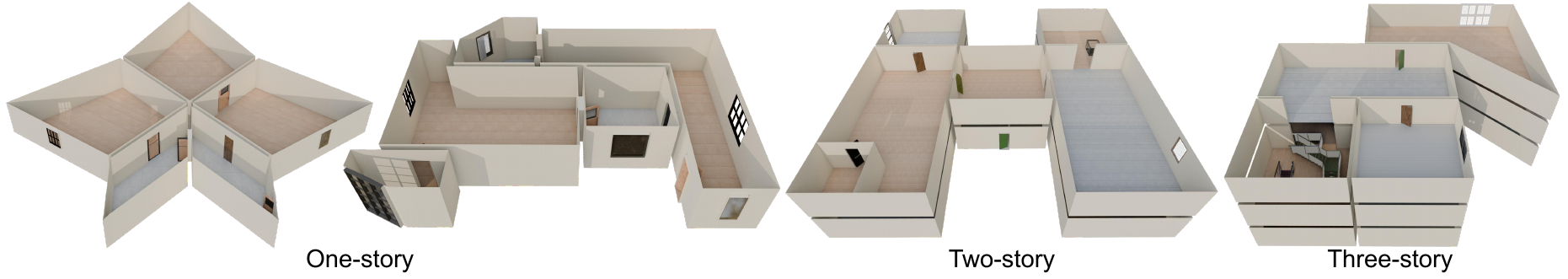}
    \caption{Visualization of diverse architectural layouts generated by Text2Villa.}
    \label{layout}
\end{figure*}

%% file: pic/human.tex
\begin{figure*}[t]
    \centering
    \includegraphics[width=\textwidth]{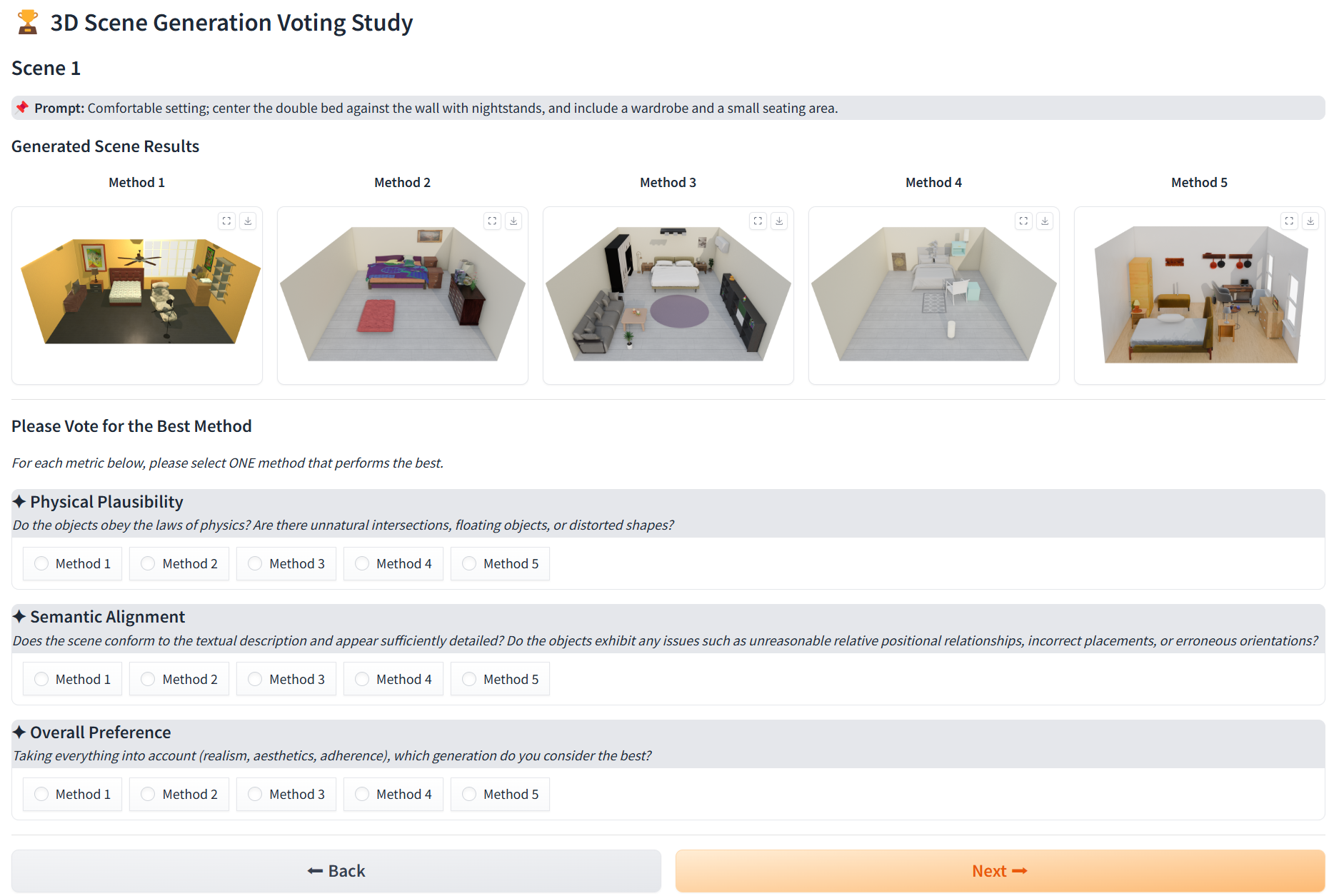}
    \caption{User interface for the human preference study.}
    \label{human}
\end{figure*}

%% file: pic/time.tex
\begin{table}[t]
    \centering
    \caption{Average single-room generation time of different methods evaluated on 27 scenes.}
    \vspace{-0.3cm}
    \label{time}
    \begin{tabular}{l c}
    \toprule
    \textbf{Method} & \textbf{Time} \\
    \midrule
    Holodeck & $\sim$5 mins \\
    MANSION & $\sim$10 mins \\
    SceneFoundry & $\sim$13 mins \\
    LayoutVLM & $\sim$13 mins \\
    SceneWeaver & $\sim$64 mins \\
    Ours & $\sim$15 mins \\
    \bottomrule
    \end{tabular}
    \vspace{-0.3cm}
\end{table}

%% file: pic/prompt2.tex
\begin{figure*}[!ht]\centering
\captionof{table}{System prompt for initializing and updating the A-PSSG.}
\begin{minipage}{\textwidth}\vspace{0mm}    \centering
\begin{tcolorbox}[llmprompt]
\small

You are an expert interior designer and a 3D scene spatial reasoning agent. Your task is to construct and update an Affordance-driven Physical-Semantic Scene Graph (A-PSSG) based on user instructions and physics engine feedback. You must strictly output in JSON format.

\textbf{Node Definitions \& Physical Affordances:}
Every object (including the \texttt{floor}) must be instantiated as a node with specific affordance attributes. A node can assume one or multiple of the following roles:
\begin{itemize}
    \item \textbf{$V_{pro}$} (Support Provider): Objects that expose valid support surfaces (e.g., floors, tables) or provide collision-free containment cavities (e.g., cabinets, bookshelves).
    \item \textbf{$V_{con}$} (Support Consumer): Objects constrained by gravity or attachment (e.g., books, desk lamps, TVs). They MUST establish a physical connection with a $V_{pro}$.
    \item \textbf{$V_{int}$} (Functional Interactor): Objects that require an interaction reservation zone along their functional orientation (e.g., sofas, wardrobes, TVs) for human activities.
\end{itemize}

\textbf{Edge Constraints:}
\begin{itemize}
    \item \textbf{$E_{sem}$} (Semantic Relation Edges): Defines relative orientation and spatial layout between objects (e.g., \texttt{"face\_to"}, \texttt{"front\_of"}, \texttt{"behind"}, \texttt{"align"}, \texttt{"right\_of"}, \texttt{"left\_of"}).
    \item \textbf{$E_{phys}$} (Physical Interaction Edges): Must be directed from a $V_{con}$ to a $V_{pro}$. 
    \begin{itemize}
        \item \texttt{"ontop"}: The bottom bounding box of the child must be coplanar with the support surface of the parent.
        \item \texttt{"inside"}: The child must be located within an independent cavity of the parent.
    \end{itemize}
\end{itemize}

\textbf{Generation \& Common-sense Rules:}
\begin{enumerate}
    \item \textbf{Explicit Extraction}: Precisely instantiate all objects explicitly mentioned in the input text.
    \item \textbf{Common-sense Supplement}: If the user input is brief, you MUST leverage common-sense priors to supplement reasonable asset categories and quantities according to the room type.
    \item \textbf{Quantity Constraint}: The total number of object nodes in the room MUST be strictly between \textbf{15 and 35} to ensure scene richness.
    \item \textbf{No Floating}: Every $V_{con}$ object must have exactly one $E_{phys}$ edge connecting it to a $V_{pro}$ object (or the floor).
    \item \textbf{Dynamic Updating (Closed-loop)}: If provided with penalty feedback (Collision/Floating/Semantic error), update the graph by adding (\texttt{add\_entity}), removing (\texttt{remove\_entity}), or modifying edge relations to resolve the conflict.
\end{enumerate}

\textbf{Task:} Analyze the given description, and generate the A-PSSG JSON based on the room type and state evaluation feedback (if any). Return ONLY the JSON format as shown below, without Markdown code blocks:
\vspace{-2mm}
{\scriptsize
\begin{verbatim}
{
  "nodes": [
    {"id": "floor_1", "category": "Floor", "affordances": ["V_pro"]},
    {"id": "shelf_1", "category": "Shelf", "affordances": ["V_con", "V_pro", "V_int"]},
    {"id": "shelf_2", "category": "Shelf", "affordances": ["V_con", "V_pro", "V_int"]},
    {"id": "tv_stand_1", "category": "TVStand", "affordances": ["V_con", "V_pro", "V_int"]},
    {"id": "tv_1", "category": "TV", "affordances": ["V_con", "V_int"]},
    {"id": "book_1", "category": "Book", "affordances": ["V_con"]},
    {"id": "armchair_1", "category": "Armchair", "affordances": ["V_con", "V_pro", "V_int"]},
    {"id": "sofa_1", "category": "Sofa", "affordances": ["V_con", "V_pro", "V_int"]},
    {"id": "table_1", "category": "Table", "affordances": ["V_con", "V_pro", "V_int"]}
  ],
  "edges_sem": [
    {"source": "shelf_2", "relation": "align", "target": "shelf_1"},
    {"source": "shelf_2", "relation": "right of", "target": "tv_stand_1"},
    {"source": "tv_stand_1", "relation": "behind", "target": "table_1"},
    {"source": "tv_1", "relation": "behind", "target": "table_1"},
    {"source": "armchair_1", "relation": "behind", "target": "table_1"},
    {"source": "table_1", "relation": "front of", "target": "sofa_1"},
    {"source": "tv_stand_1", "relation": "face to", "target": "sofa_1"},
    {"source": "tv_1", "relation": "face to", "target": "sofa_1"},
    {"source": "table_1", "relation": "face to", "target": "sofa_1"}
  ],
  "edges_phys": [
    {"source": "shelf_1", "relation": "ontop", "target": "floor_1"},
    {"source": "shelf_2", "relation": "ontop", "target": "floor_1"},
    {"source": "tv_stand_1", "relation": "ontop", "target": "floor_1"},
    {"source": "sofa_1", "relation": "ontop", "target": "floor_1"},
    {"source": "table_1", "relation": "ontop", "target": "floor_1"},
    {"source": "armchair_1", "relation": "ontop", "target": "floor_1"},
    {"source": "tv_1", "relation": "ontop", "target": "tv_stand_1"},
    {"source": "book_1", "relation": "inside", "target": "shelf_1"}
  ]
}
\end{verbatim}
}

\end{tcolorbox}
\end{minipage}
\label{prompt2}
\end{figure*}